%% file: main.tex
\documentclass[journal]{IEEEtran}
\usepackage{graphicx}
\usepackage{cite}
\usepackage{picinpar}
\usepackage{amsmath}
\usepackage{url}
\usepackage{flushend}
\usepackage[latin1]{inputenc}
\usepackage{colortbl}
\usepackage{soul}
\usepackage{multirow}
\usepackage{pifont}
\usepackage{color}
\usepackage{alltt}
\usepackage[hidelinks]{hyperref}
\usepackage{enumerate}
\usepackage{breakurl}
\usepackage{epstopdf}
\usepackage{pbox}

\usepackage{fancyhdr}
\usepackage{lineno} 
\usepackage{bm}

\usepackage{amssymb}
\usepackage{stfloats}
\usepackage{algorithmic}
\usepackage[ruled,vlined]{algorithm2e}

\usepackage[table]{xcolor}
\usepackage{comment}
\newcommand{\hw}[1]{\textcolor{black}{#1}} 
\begin{document}

\title{Privacy-Preserving Transactive Energy Management for IoT-aided Smart Homes via Blockchain}

\author{Qing~Yang,~\IEEEmembership{Member,~IEEE,}
        Hao~Wang,~\IEEEmembership{Member,~IEEE}
\thanks{This work is in part supported by the National Natural Science Foundation of China (project 61901280) and the FIT Academic Staff Funding of Monash University. (Corresponding author: Hao Wang.)}
\thanks{Q. Yang is with the Blockchain Technology Research Center (BTRC) and the College of Electronics and Information Engineering (CEIE), Shenzhen University, Shenzhen, Guangdong Province, PRC, e-mail: {yang.qing@szu.edu.cn}.}
\thanks{H. Wang is with the Department of Data Science and Artificial Intelligence, Faculty of Information Technology, Monash University, Melbourne, VIC 3800, Australia, and Stanford Sustainable Systems Lab, Stanford University, CA 94305, USA, e-mail: hao.wang2@monash.edu.}
}

\maketitle

\begin{abstract}
With the booming of smart grid, The ubiquitously deployed smart meters constitutes an energy internet of things. This paper develops a novel blockchain-based transactive energy management system for IoT-aided smart homes. We consider a holistic set of options for smart homes to participate in transactive energy. Smart homes can interact with the grid to perform vertical transactions, e.g., feeding in extra solar energy to the grid and providing demand response service to alleviate the grid load. Smart homes can also interact with peer users to perform horizontal transactions, e.g., peer-to-peer energy trading. However, conventional transactive energy management method suffers from the drawbacks of low efficiency, privacy leakage, and single-point failure. To address these challenges, we develop a privacy-preserving distributed algorithm that enables users to optimally manage their energy usages in parallel via the smart contract on the blockchain. Further, we design an efficient blockchain system tailored for IoT devices and develop the smart contract to support the holistic transactive energy management system. Finally, we evaluate the feasibility and performance of the blockchain-based transactive energy management system through extensive simulations and experiments. The results show that the blockchain-based transactive energy management system is feasible on practical IoT devices and reduces the overall cost by 25\%.
\end{abstract}


\markboth{IEEE Internet of Things Journal}%
{}

\definecolor{limegreen}{rgb}{0.2, 0.8, 0.2}
\definecolor{forestgreen}{rgb}{0.13, 0.55, 0.13}
\definecolor{greenhtml}{rgb}{0.0, 0.5, 0.0}
\SetAlFnt{\small}
\SetAlCapFnt{\small}

\section{Introduction}\label{sec:intro}
\input{introduction.tex}

\section{Related Works}\label{sec:related}
\input{related.tex}

\section{System Design}\label{sec:model}
\input{model.tex}

\section{Problem Formulation and Algorithm Design}\label{sec:formulation}
\input{formulation.tex}

\section{System Implementation and Performance Evaluation}\label{sec:eval}
\input{evaluation.tex}

\section{Conclusion}\label{sec:conclusion}
This paper presented a blockchain-based transactive energy management system that allows IoT-aided smart homes to interact with the grid and peer users via a blockchain network. We considered two dimensions of transactive energy: vertical and horizontal transactions. In vertical transactions, smart homes can choose to sell PV energy to the grid and perform demand response. In horizontal transactions, smart homes can trade energy with others in need. We developed a privacy-preserving distributed algorithm to optimize the users' transactive energy management without revealing their private information. We further implemented a blockchain system on IoT devices to support the decentralized transactive energy management platform. Simulations and experiments show that the blockchain-based transactive energy management is feasible on practical IoT devices and can reduce the overall cost by 25\%.

In our future work, we will reduce the computational complexity of the decentralized algorithms in order to support large-scale IoT network that involves tens of thousands of nodes. We are building a larger testing IoT network to analyze the performance of the proposed blockchain with massive users. We will also explore methods to efficiently implement complex transactive-energy algorithms in the smart contract by WebAssembly \cite{WebAssembly} and predefined contracts.



\bibliographystyle{IEEEtran}
\bibliography{IEEEabrv,ref.bib} 

\end{document}

%% file: introduction.tex
\IEEEPARstart{S}{mart} meters, as the communication and computing modules of smart homes, are widely deployed with the booming application of the smart grid. Benefited from the development of the Internet of Things (IoT) technology such as edge computing and 5G narrowband \cite{li2017smart}, the smart meter can achieve sophisticated functions for efficient data communication with limited hardware resources and monitoring and management of electric appliances \cite{sun2015comprehensive}. These inter-connected smart meters constitute an energy IoT (EIoT) network that enables the exchange of both electrical energy and digital information in the smart grid. In this context, transactive energy \cite{moghaddam2018fog}, which enables prosumers to interact with other smart grid entities in a marketplace, emerges as a multi-disciplinary research topic that aims to facilitate a smarter EIoT system.

Recently, the application of blockchain technology in the context of IIoT (industrial IoT) and smart grids inspires efforts in both academia and industry \cite{mollah2020blockchain}. Blockchain, a disruptive technology originates in digital currency, is recently gaining momentum in various areas. Bitcoin \cite{nakamoto2008bitcoin} is the first successful application of the blockchain technology that implements a tamper-proof distributed ledger to record all the transactions. Ethereum \cite{eth} introduces the \emph{smart contract} by supporting the Ethereum Virtual Machine (EVM) on top of its blockchain. The smart contract allows people to utilize the blockchain as a trustable computing machine, thus facilitates the prosperity of decentralized applications (DApps). 

Due to its versatility and decentralization nature, the integration of blockchain brings about paradigm shifts in many industries, including energy trading and transactive energy \cite{dai2019blockchain}. Many existing studies have focused on various aspects of the blockchain system in IIoT and smart grids, including privacy protection mechanisms and blockchain-based energy trading systems in \cite{gai2019privacy, li2017consortium, wan2019blockchain, mollah2020blockchain, li2019blockchain, gai2019permissioned, yang2020blockchain}. The review and comparison of these related studies are presented in Section~\ref{sec:related}. Our literature review finds that it is necessary to consider the following issues to implement a feasible blockchain-empowered transactive energy system in the energy IoT environment. 1) Can the blockchain-empowered transactive energy system be intelligent to maximize the efficiency of the grid? 2) Can the blockchain-empowered transactive energy system preserve the users' privacy information, including identity and energy supply/demand record? 3) Is the proposed blockchain solution implementable on IoT devices such as smart meters?

To address the challenges as mentioned above, we present a privacy-preserving transactive energy management system based on IoT blockchain. 
We developed a blockchain-based transactive energy system for smart homes. The smart homes can participate in a set of holistic transactive-energy options. For example, in vertical transactions with the operator, smart homes can sell extra PV energy to the grid and provide demand response (DR) service to the grid. For horizontal transactions with peer smart homes, smart homes can trade energy with other smart homes in the community to gain benefits. We address the three challenges of privacy, efficiency, and implementation posed on the EIoT system and summarize the main contributions as follows.
\begin{enumerate}[1)]
  \item \textit{Efficient transactive energy system:} We develop a holistic transactive energy system that enables smart homes to interact with the grid and other peer users in the energy IoT system. We demonstrate the benefits of transactive energy to smart homes for reducing their energy costs and to the system for facilitating feed-in PV energy and demand response.
  \item \textit{Privacy-preserving transactive energy management:} We design a distributed algorithm for transactive energy management that consists of each user's transactive-energy decision making with the smart contract and preserves users' privacy.
  \item \textit{Validated Blockchain implementation on IoT devices:} We implement and validate the blockchain-based transactive energy platform on IoT devices that have limited hardware resources. The experiments on IoT devices demonstrate the effectiveness of our design of the blockchain system and smart contract.
\end{enumerate}

The remainder of the paper is organized as follows. Section~\ref{sec:related} introduces the background and related works. Section~\ref{sec:model} describes the system model of the blockchain-based transactive energy platform. Section~\ref{sec:formulation} formulates the energy trading problem and presents the distributed transactive energy algorithm on a blockchain. Section~\ref{sec:eval} evaluates the proposed system with extensive experiments and simulations. Section~\ref{sec:conclusion} concludes this paper.

%% file: related.tex
This section introduces some existing works related to blockchain-based transactive energy management in the EIoT environment such as smart grid, smart city, and industrial IoT. When reviewing the literature, we focus on three particular aspects: 1) user privacy protection mechanism; 2) transactive energy management algorithm and its performance; 3) design of the underlying blockchain system. Table~\ref{t1:related} summarizes the differences between the related works and our work.

\begin{table*}[b]
\centering
\caption{The comparison of the related existing works and this work.}\label{t1:related}
  \begin{tabular}{p{0.7cm}|p{1.2cm}|p{7cm}|p{7cm}}
    \hline
    \textbf{Topic} & \textbf{Paper} & \textbf{Existing Method}  & \textbf{Our Method}\\
    \hline
    \multirow{3}{0.7cm}{\rotatebox[origin=c]{90}{\parbox[c]{2cm}{\centering Privacy protection}}} & \cite{aitzhan2016security} & Used the multi-signature algorithm to hide the users' private information at the cost of slow data processing and bloated transaction size. & Use the elliptic curve digital signature and distributed optimization algorithm for faster data processing and smaller transaction size. \\
    \cline{2-4}
    & \cite{gai2019privacy} & Employed a private address-account mapping method to hide the user's identity. & \multirow{2}{7cm}{$\bullet$ Proposed a distributed P2P energy management algorithm that does not reveal the users' private information.\\
    $\bullet$ Remove the need of the central identity management node.} \\
    \cline{2-3}
    & \cite{li2017consortium} & Used a trusted identity management node and a encrypted account pool to hide the clients' identity. &\\
    \cline{2-4}
    & \cite{wan2019blockchain} & Designed an access control mechanism to prevent privacy leakage. & Employ the open-access consortium blockchain and smart contract. \\
    \hline
   \multirow{3}{0.7cm}{\rotatebox[origin=c]{90}{\parbox[c]{2cm}{\centering Energy management}}} & \cite{mihaylov2014nrgcoin,exergy} & Blockchain projects with built-in token to facilitate the electric payment of P2P energy trading. & Use blockchain and smart contract for decentralized energy management and payment. \\
    \cline{2-4}
    & \cite{wang2019energy} & Designed a crowdsourcing-based P2P energy trading algorithm and tested the algorithm on Hyperledger blockchain. &  \multirow{4}{7cm}{$\bullet$ Design a distributed P2P energy trading algorithm based on the ADMM method.\\
    $\bullet$ Consider a holistic energy management for various appliances and privacy protection. \\
    $\bullet$ Validate the system on the blockchain tailored for IoT devices.
    } \\
    \cline{2-3}
    & \cite{sabounchi2017towards} & Proposed an action-based trading algorithm on Ethereum. &  \\
    \cline{2-3}
    & \cite{li2017consortium} & Proposed a blockchain-based credit system for secure energy trading. &  \\
    \cline{2-3}
    &  \cite{wang2019bbars} & Blockchain-based rewarding scheme for vehicle battery management. &  \\
    \hline
    \multirow{3}{0.7cm}{\rotatebox[origin=c]{90}{\parbox[c]{2cm}{\centering Blockchain Infrastructure}}} & \cite{li2017consortium} & Proposed a simplified consensus protocol to reduce the computational complexity of the consensus protocol. &  \multirow{4}{7cm}{$\bullet$ Build a high-performance IoT blockchain based on Quorum.\\
    $\bullet$ Improve the PBFT consensus protocol to adapt the hardware of IoT device. \\
    $\bullet$ Validate and test the blockchain system on a practical IoT network.} \\
    \cline{2-3}
    & \cite{gai2019privacy,wan2019blockchain} & Used the Hyperledger Fabric blockchain, which is infeasible on IoT devices. &  \\
    \cline{2-3}
    & \cite{thomas2019general} & Implemented by smart contract on the Ethereum blockchain.&  \\
    \cline{2-3}
    & \cite{iota, danzi2019delay} & IOTA is a DAG-based blockchain for IoT applications, but with low throughput and long transaction delay. &  \\
    \hline
  \end{tabular}
\end{table*}

With the wide deployment of IoT devices, leakage of privacy becomes a vital concern for many IoT applications, including smart grid and transactive energy \cite{zheng2018user, cha2018privacy, pnnl}. Although blockchain adopts pseudonymity to conceal the users' real identity based on asymmetric cryptography, the publicity of the block data threatens users' privacy \cite{zhang2019security, dorri2017blockchain}. Aitzhan \textit{et al.} \cite{aitzhan2016security} firstly employed the multi-signature algorithm to secure the users' privacy during energy trading, but at the cost of slow data processing and bloated transaction size. To address the privacy issue in smart grid, Gai \textit{et al.} designed a private address-account mapping method in \cite{gai2019privacy}, and Li \textit{et al.} proposed to use a encrypted account pool to hide the clients' identity in \cite{li2017consortium}. However, the effectiveness of these methods relied on a trustable identity management authority, which limited their application in permissionless scenarios. In \cite{wan2019blockchain}, the authors developed an access-control algorithm to strengthen the security and privacy for blockchain-based IIoT systems. However, \cite{wan2019blockchain} works only on a private blockchain and thus cannot scale well. In this work, instead of trying to secure the private information transmitted in the blockchain, we implement the transactive energy management algorithm in a distributed manner based on IoT blockchain. The proposed method allows users to process private information locally without revealing it on the blockchain, thus effectively preserving their privacy.

As a disruptive technology, blockchain has recently inspired a lot of paradigm shifts in both academic and industrial areas of transactive energy \cite{mollah2020blockchain, li2019blockchain, gai2019permissioned}. The industry has adopted blockchain as a convenient energy-sharing platform as well as a secure payment tool. NRGCoin \cite{mihaylov2014nrgcoin} was initiated as a digital currency dedicated to renewable energy trading. Exergy \cite{exergy} developed a blockchain for the trading of distributed energy resources (DER) in IoT scenarios and deployed it on a microgrid in Brooklyn of New York City. In academic research, the blockchain has been employed to improve the efficiency of transactive energy systems. Wang \textit{et al.} proposed a P2P (peer-to-peer) energy crowdsourcing algorithm and tested it on Hyperledger in \cite{wang2019energy}. Sabounchi \textit{et al.} \cite{sabounchi2017towards} used the Ethereum smart contract to implement a transactive energy trading algorithm based on auction theory. In both \cite{wang2019energy} and \cite{sabounchi2017towards}, the users' private information, including power consumption records and trading prices, are disclosed on the blockchain. To address the privacy issue, Li \textit{et al.} \cite{li2017consortium} designed a blockchain-based credit system to guarantee the privacy and security of the proposed transactive energy trading platform in IIoT. Unlike the centralized energy management algorithm used by \cite{li2017consortium}, our transactive energy management algorithm in this paper is distributed without any central control. In \cite{wang2019bbars}, the authors design a blockchain-based rewarding scheme for the vehicle-to-grid (V2G) system to incentivize energy exchange. Compared with \cite{wang2019bbars}, our work considers a holistic option of transactive energy, including demand response, feed-in PV energy, and P2P energy trading, to optimize the efficiency of the whole grid.

Blockchain, initially introduced in Bitcoin \cite{nakamoto2008bitcoin} as a secure, tamper-proof, and verifiable database, has been extensively used in various IoT systems \cite{dai2019blockchain, ferrag2018blockchain}. However, in IoT applications, limited hardware resources, including storage, network bandwidth, and computing power, pose unique challenges to the blockchain \cite{ferrag2018blockchain}. The consensus protocol is the mechanism used in blockchain to synchronize all the distributed nodes. In \cite{li2017consortium}, the authors designed a simplified consensus protocol to reduce the computational complexity for IoT devices, but such simplification hurdles the liveness of the consensus protocol. Both \cite{gai2019privacy} and \cite{wan2019blockchain} adopted the Hyperledger Fabric blockchain and conducted experiments on PCs with Intel CPUs, but their setup is infeasible on IoT devices without high-performance CPUs. In \cite{thomas2019general}, Lee \textit{et al.} proposed a smart-contract-based shared control mechanism for energy system and implemented it with Solidity on Ethereum \cite{eth}. However, the mining algorithm of Ethereum consumes exorbitant amounts of power and memory resources that are unaffordable for IoT devices such as smart meters. IOTA, a blockchain project targeting IoT applications, adopts the directed acyclic graph (DAG) structure and uses a new consensus protocol (the Tangle) to allow IoT devices to join the mining process \cite{iota}. Nevertheless, the low throughput and long transaction confirmation delay degrade the performance of blockchain-based IoT applications \cite{danzi2019delay}. In this paper, we tailor the transactive energy management algorithm and the blockchain design for IoT devices and test its feasibility on a practical IoT network.

%% file: model.tex
We present the architecture of the blockchain-based transactive energy management system on the IoT-aided smart meters. We first present the smart home's model, including notations and models for loads, generations, and electric vehicles. We then introduce the principle of the transactive energy management system consisting of the feed-in tariff, demand response, and energy trading. Finally, we elaborate on the design of the blockchain for IoT-aided smart homes.

\begin{figure}[!t]
    \centering
    \includegraphics[width=8.0cm]{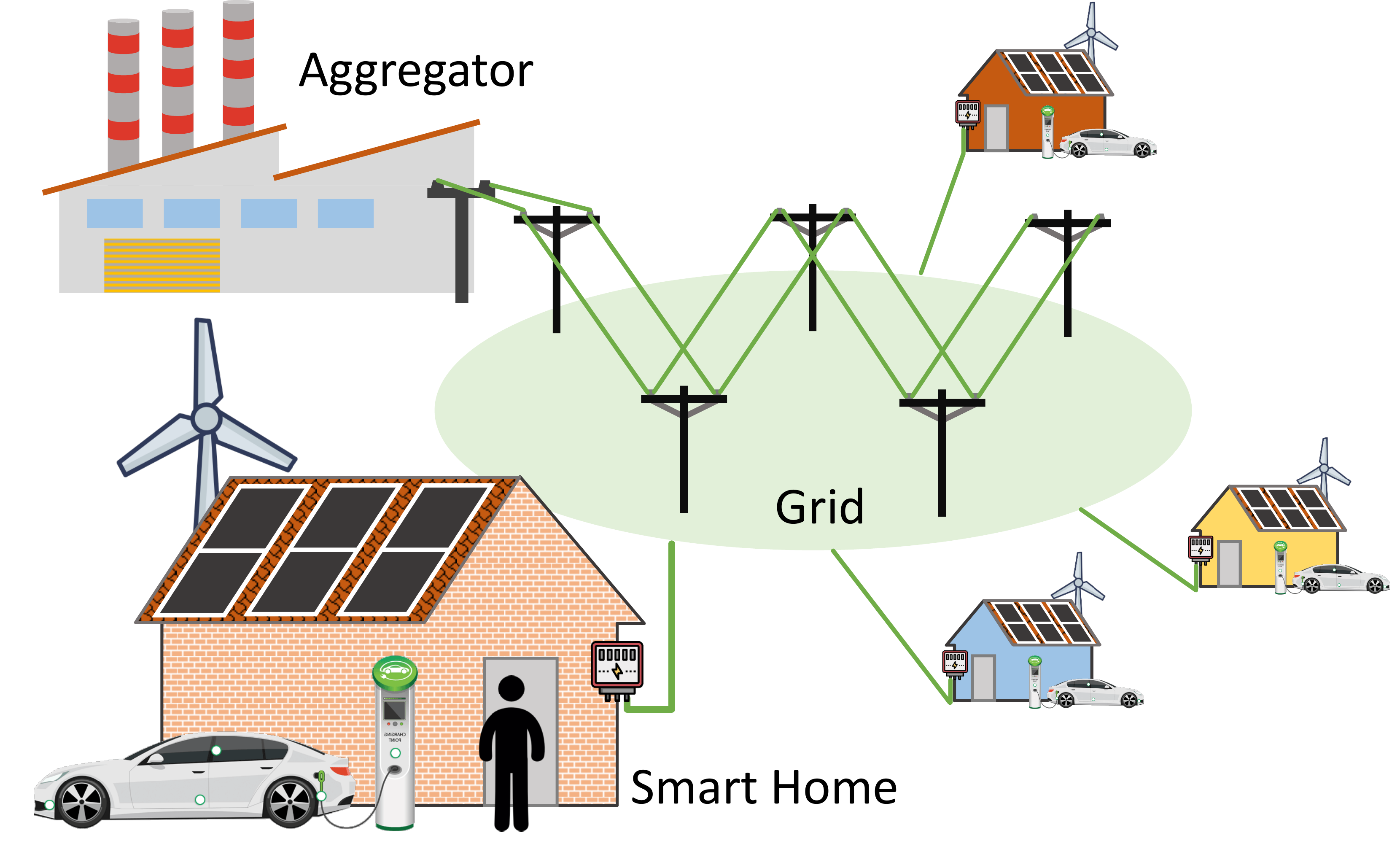}
    \caption{The system model of the blockchain-based transactive energy management system.}
    \label{f1:sysmod}
\end{figure}

\subsection{The Smart Home Model}
With the ubiquitous deployment of IoT devices, the smart home can intelligently manage various electrical appliances. As shown in Fig.~\ref{f1:sysmod}, we consider 1) electric load that consumes energy to sustain the smart home, e.g., refrigerators, air conditioners, washers, coffee machines; 2) local renewable energy generations from photovoltaics (PV) panels and wind turbines; 3) electric vehicles (EVs) that can charge, discharge, and store energy. The smart home connects to the power grid and interacts with other smart homes and the aggregator. The aggregator is the operator of the local grid as well as the proposed energy management platform. 

As the ``eyes'' and ``brain'' of smart homes, the smart meter is an intelligent IoT device that automatically manages the supply and load; furthermore, the smart meter handles the exchange of energy and information between the smart home and other parts of the grid. We define the owner of the smart home as the \emph{user} of the transactive energy management system and let $n \in \mathcal{N} = \{1,2,\dots,N \}$ denotes all the users in the system. The transactive energy management works in a time-slotted manner, i.e., $t \in \mathcal{T} = \{1,2,\dots,H\}$ denotes the number of time slot and $T$ denotes the maximum scheduling window. 

\subsubsection{The Electric Loads of the Smart Home}
User $n$'s smart home appliances can be classified into the following types. The first type of load is shiftable over time, such as the dryer and washer. The second type of load is curtailable, and for example, entertainment and recreation activities using pool pumps can be curtailed. The third type of appliances is the adjustable load, such as heating, ventilation, and air conditioning (HVAC) units. The rest of the load is inflexible and uninterruptible as it is used to meet the essential needs, e.g., light and refrigerator. We present the model for each type of load as follows.

The shiftable load $l_n^{\text{S}}[t]$ represents the appliances that the user $n$ can allocate over a shiftable time window $\mathcal{T}_n^{\text{S}}$. For example, the smart grid can automatically schedule the washer to work at any proper time slot within the available time window. However, each user has its preferred load profile for the shiftable load, which is denoted by $L_n^{\text{S}}[t]$. To complete the load profile within the scheduling window, The scheduled shiftable load must satisfy the following constraint
    \begin{align}
                \sum_{t} l_n^{\text{S}}[t] = \sum_{t} L_n^{\text{S}}[t], \, t \in \mathcal{T}_n^{\text{S}}, ~n \in \mathcal{N}. \label{constraint-load3}
    \end{align}

Note that when users choose to shift the load from the preferred load profile, they change their routine behavior and experience discomfort. We explicitly model the cost of load shifting as
    \begin{align}
            C_n^{\text{S}} = \omega_{\text{S}} \sum_{t} \left( l_n^{\text{S}}[t] - L_n^{\text{S}}[t] \right)^{2}, \, t \in \mathcal{T}_n^{\text{S}}, \label{objective-load2}
    \end{align}
where the coefficient $\omega_{\text{S}}$ denotes users' sensitivity of the behavior change due to shifted load. Note that the quadratic cost is widely used in energy economics literature to characterize the users' marginal discomfort that often becomes severe as deviation enlarges. 

The second type of smart home load is the curtailable load denoted by $l_n^{\text{C}}[t]$ for user $n$. We let $L_n^{\text{C}}[t]$ denote user $n$'s originally planned load. The user can curtail this load at different levels to trade off their needs against the costs. Specifically, user $n$ schedules its curtailable load $l_n^{\text{C}}[t]$ that satisfies
\begin{align}
    0 \leq l_n^{\text{C}}[t] \leq L_n^{\text{C}}[t], ~n \in \mathcal{N}. \label{constraint-load5}
\end{align}
Similarly, when users choose to curtail the load, they sacrifice some of their planned activities, and the cost is also modeled as a function of the curtailed load, i.e.,     \begin{align}
            C_n^{\text{C}} = \omega_{\text{C}} \sum_{t} \left( l_n^{\text{C}}[t] - L_n^{\text{C}}[t] \right)^{2}, \label{objective-load3}
    \end{align}
where $\omega_{\text{C}}$ is the sensitivity coefficient of user $n$ on its curtailed load.

The third type of smart home load is the adjustable load (AL), for which we focus our analysis on the HVAC load. The HVAC system consumes electricity power $l_n^{\text{A}}[t]$ to control the indoor temperature at $\text{Tin}^n[t]$ in time slot $t$. The dynamics of the indoor temperature \cite{cui2019} follows
    \begin{align}
            \text{Tin}^n[t] = \text{Tin}^n[t-1] + \alpha l_n^{\text{A}}[t] - \beta (\text{Tin}^n[t-1] - \text{Tout}^n[t]), \label{constraint-load1}
    \end{align}
where $\text{Tout}^n[t]$ denotes the outdoor temperature. Coefficients $\alpha$ and $\beta$ are the HVAC parameters indicating the working efficiency and mode. The sign of $\beta$ indicates the HVAC's working modes, specifically, positive for cooling and negative for heating.

Users often have setpoint temperature $\text{Tref}^n[t]$ for the HVAC, and any deviation from the setpoint will cause discomfort to users. We measure the discomfort by the difference between the indoor temperature and its setpoint for user $n$ as
    \begin{equation}
            C_n^{\text{A}}=  \omega_{\text{A}} \sum_{t} \left( \text{Tin}^n[t] - \text{Tref}^n[t] \right)^{2}, ~t \in \mathcal{T}, \label{objective-load1}
    \end{equation}
where $\omega_{\text{A}}$ denotes the user's sensitivity coefficient to the indoor temperature. Note that the users experience greater discomfort when indoor temperature deviates more. The indoor temperature should be also controlled within a range $[ \underline{\text{Tin}}^n, \overline{\text{Tin}}^n ]$, where $\underline{\text{Tin}}^n$ and $\overline{\text{Tin}}^n$ are the lower-bound and upper-bound of the tolerable indoor temperature of user $n$. 

The rest of the smart home load is the inflexible load denoted by $l_n^{\text{I}}[t]$ in time slot $t$. Different from the adjustable load $l_n^{\text{A}}[t]$, the shiftable load $l_n^{\text{S}}[t]$, and the curtailable load $l_n^{\text{C}}[t]$, the user cannot control its inflexible load $l_n^{\text{I}}[t]$.

\subsubsection{The Power Supply Models}
The smart home has electricity supply from two sources: first, user $n$ can purchase electricity from the grid denoted by $s_n^{\text{G}}[t]$; and second, user $n$ can use its renewable energy denoted by $s_n^{\text{R}}[t]$. Users can even trade surplus energy with other parties of the grid, which will be discussed later in this section. 

Note that the grid power and renewable power are upper-bounded by $S_{\text{G}}$ and $S_n^{\text{R}}[t]$, which denote the maximum powerline capacity and the available renewable generation, respectively. We assume that the maximum powerline $S_{\text{G}}$ is the same for all smart homes in the grid. The renewable generation $S_n^{\text{R}}[t]$, however, depends on the solar and wind condition of each smart home.

To incentivize peak shaving for the grid, the pricing strategy of the grid consists of a regular usage price $p_{\text{G}}$ and a peak usage price $p^{\star}_{\text{G}}$. Specifically, user $n$ needs to pay
    \begin{align}
            C_n^{\text{G}} = p_{\text{G}} \sum_{t} s_n^{\text{G}}[t] + p^{\star}_{\text{G}} \max_{t} s_n^{\text{G}}[t], ~t \in \mathcal{T}, \label{objective-supply} 
    \end{align}
where $p_{\text{G}} \sum_{t} s_n^{\text{G}}[t]$ is the bill for total electricity usage and $p^{\star}_{\text{G}} \max_{t} s_n^{\text{G}}[t]$ is the bill for peak usage.

\subsubsection{The Model of the Electric Vehicle (EV)} 
We seperate the discussion of EV from the load model, as EV can perform vehicle-to-home (V2H) to discharge battery. As shown in Fig.~\ref{f1:sysmod}, we assume that user $n$ has an EV parked at home during the period $\mathcal{T}_n^{\text{V}} \triangleq [t_n^{\text{A}},t_n^{\text{D}}]$, where $t_n^{\text{A}}$ denotes the arrival time  and $t_n^{\text{D}}$ denotes departure time. The EV needs to be fully charged before departure, and can also be discharged to serve the household load during $\mathcal{T}_n^{\text{V}}$ at certain cost of battery degradation. We denote $e_n^{\text{V}}[t]$ as user $n$'s EV battery energy level, and $E_n^{\text{V}}$ as its battery capacity such that $e_n^{\text{V}}[t] \in [0, E_n^{\text{V}}]$. Furthermore, the EV battery should be charged to meet the need of travel before the departure time, i.e., 
\begin{align}
     e_n^{\text{V}}[t_n^{\text{D}}] = E_n^{\text{V}}, ~n \in \mathcal{N}. \label{constraint-ev2}
\end{align}

We denote $p_n^{\text{cha}}[t]$ as the charging power and $p_n^{\text{dis}}[t]$ as the discharging power in time slot $t$. We bound them by $p_n^{\text{cha}}[t] {\in} [0, P_n^{\text{cha}}]$ and $p_n^{\text{dis}}[t] {\in} [0, P_n^{\text{dis}}]$, where $P_n^{\text{cha}}$ and $P_n^{\text{dis}}$ are the charging and discharging limits for the EV's battery of user $n$, respectively.

The energy stored in the battery varies over time, according to the charging and discharging operation. Its dynamics follows
\begin{align}
   e_n^{\text{V}}[t] {=} e_n^{\text{V}}[t{-}1] {+} \mu_n p_n^{\text{cha}}[t] {-} p_n^{\text{dis}}[t] / \nu_n , \label{constraint-ev5}    
\end{align}
where the two parameters $\mu_n {\in} [0,1]$ and $\nu_n {\in} [0,1]$ denote the charging and discharging efficiency of user $n$'s EV battery, respectively. 

During the period $\mathcal{T}_n^{\text{V}}$, user $n$ can perform V2H to discharge the EV battery when it's needed to serve the household load. However, doing so will incur battery degradation, and we model the EV battery degradation cost as
    \begin{align}
            C_n^{\text{V}} = \omega_{\text{V}} \sum\nolimits_{t\in \mathcal{T}_n^{\text{V}}} \left(p_n^{\text{dis}}[t]\right)^2, \label{objective-battery}
    \end{align}
where $\omega_{\text{V}}$ is the cost coefficient, and the quadratic form reflects a more significant degradation when the EV battery is discharged more deeply. Note that we do not consider the battery degradation when the EV is used outside the home.

\subsection{The Transactive Energy Management Model}\label{subsec:transactive}
\begin{figure}[!t]
    \centering
    \includegraphics[width=8.7cm]{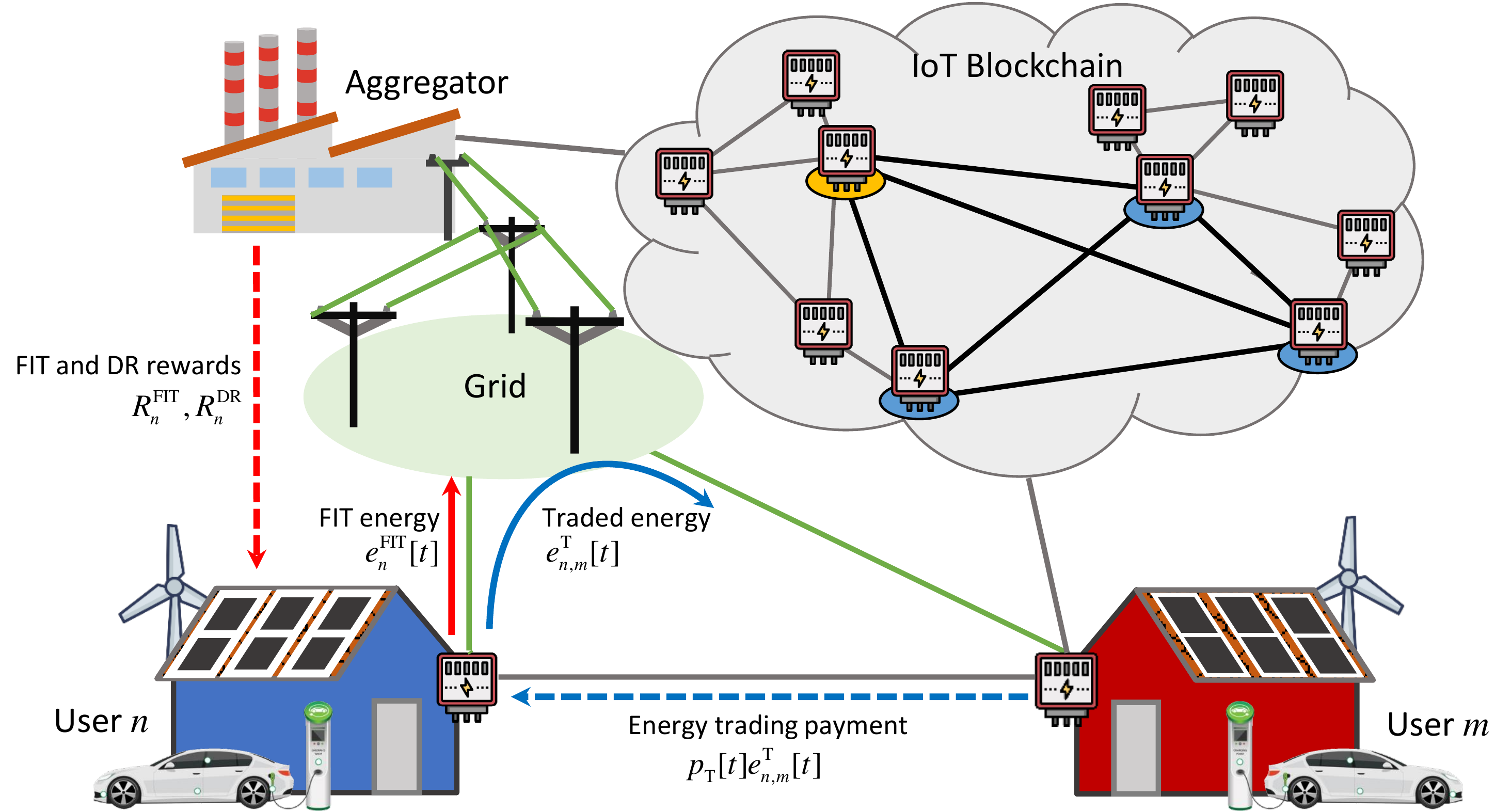}
    \caption{The system design of the blockchain-based transactive energy management system.}%
    \label{f2:sysdesign}
\end{figure}

This section focuses on the transactive energy management that enables users to interact with other parties in the grid via the IoT blockchain as shown in Fig.~\ref{f2:sysdesign}. We define two types of energy transactions, namely \emph{vertical transaction} and \emph{horizontal transaction}, according to the role of the trading counterparty. The vertical transactions include feed-in tariff (FIT) transactions and demand response. With the FIT transaction, users can sell PV generation to the grid and earn the feed-in tariff. In the event of demand response, users can reduce originally scheduled demand to alleviate peak in the system through aggregators. For the horizontal transactions, users can trade energy with peer users to leverage their diverse patterns in using EV, HVAC, and scheduling different types of load.

\subsubsection{Energy Trading of the Vertical Transactions} As a policy tool designed to increase the adoption of renewable energy technologies (e.g., PV installation), FIT  programs have been widely implemented. A typical FIT program provides momentary payments to FIT-eligible renewable generators for the feed-in electricity to the grid. We assume that the utility sets a FIT denoted as $p_{\text{FIT}}[t]$ for all the users. The user $n$ choose to sell $e_n^{\text{FIT}}[t]$ from their renewable generation to the grid, such that 
\begin{align}
    0 \leq e_n^{\text{FIT}}[t] \leq S_n^{\text{R}}[t], &~ \forall n \in \mathcal{N}, \forall t \in \mathcal{T}, \label{constraint-load8} \\
    s_n^{\text{R}}[t] + e_n^{\text{FIT}}[t] \leq S_n^{\text{R}}[t], &~ \forall n \in \mathcal{N}, \forall t \in \mathcal{T}, \label{constraint-load9}
\end{align}
where the feed-in renewable energy $e_n^{\text{FIT}}[t]$ is non-negative and bounded by the available renewable generation $S_n^{\text{R}}[t]$. Also, the sum of renewable energy that supplies local demand $s_n^{\text{R}}[t]$ and that sold to the grid $e_n^{\text{FIT}}[t]$ should be no greater than the available renewable generation $S_n^{\text{R}}[t]$.

For the feed-in renewable energy, user $n$ can get a FIT reward as
\begin{align}
   R_n^{\text{FIT}} = \sum\nolimits_{t\in \mathcal{T}} p_{\text{FIT}}[t] e_n^{\text{FIT}}[t]. \label{objective-fit} 
\end{align}

Another vertical transaction is the demand response (DR), which is used by the utility or the aggregator to signal the users for load reduction in a time window $\mathcal{T}_{\text{DR}}$ (which is usually late afternoon and evening). During this window, users can choose to respond to the DR signals by reducing their grid load and then earn rewards. If user $n$ responds to the DR signals and reduce their load from scheduled grid load by $e_n^{\text{DR}}[t]$, this reduction is rewarded by a unit price $p_{\text{DR}}[t]$. The load reduction satisfies the following constraint
\begin{align}
    0 \leq e_n^{\text{DR}}[t] \leq s_n^{\text{G}}[t], &~ \forall n \in \mathcal{N}, \forall t \in \mathcal{T}_{\text{DR}}, \label{constraint-load10} 
\end{align}
which limits the load reduction to be non-negative and bounded by the scheduled grid load $s_n^{\text{G}}[t]$. Therefore, by responding the DR, user $n$ can get a reward from the grid aggregator
\begin{align}
   R_n^{\text{DR}} = \sum\nolimits_{t\in \mathcal{T}_{\text{DR}}} p_{\text{DR}}[t] e_n^{\text{DR}}[t]. \label{objective-dr} 
\end{align}

\subsubsection{Energy Trading of the Horizontal Transactions} For horizontal transactions, user $n$ can form trading pairs with user $m \in \mathcal{N} \backslash n$ to exchange energy of amount $e_{n,m}^{\text{T}}[t]$. Note that $e_{n,m}^{\text{T}}[t] >0$ if user $n$ sells energy to user $m$ in time slot $t$; otherwise, $e_{n,m}^{\text{T}}[t] <0$ if user $n$ purchases energy from user $m$. Since users are located close to each other, we assume that the loss of energy during the exchange is negligible. Therefore, we have the following clearing constraints for the horizontal transaction
\begin{align}
    e_{n,m}^{\text{T}}[t] + e_{m,n}^{\text{T}}[t] = 0,&~\forall t \in \mathcal{T},~\forall n \in \mathcal{N},~\forall m \in \mathcal{N} \backslash n. \label{constraint-load11}
\end{align}

The energy-trading partners make transactions based on the transactive energy prices $p_{\text{T}}[t]$ sent by the distribution system.\footnote{We focus on the transactive energy management of smart homes and the design of the algorithm and blockchain system. The role of system operators, e.g., optimizing transactive energy prices, is beyond the scope of this work and we consider it as future work.} 
Therefore, users who sell energy will get payments from their counterparts at prices $p_{\text{T}}[t]$. Similarly, users pay their counterparts if they purchase energy. User $n$'s reward in energy trading is 
    \begin{align}
        R_n^{\text{T}} = \sum_{t \in \mathcal{T}} \left( p_{\text{T}}[t] \sum_{m \in \mathcal{N} \backslash n} e_{n,m}^{\text{T}}[t] \right). \label{objective-et}
    \end{align}

Transactive energy management focuses on the distributed algorithmic design to facilitate both vertical transactions and horizontal transactions. We will model the transactive energy management problem and develop a distributed algorithm in Section \ref{sec:formulation}. 
\subsection{The IoT Blockchain on the Smart Meters}\label{sec:blockchainlayer}

This section elaborates on the design of the IoT blockchain that runs on smart meters, as plotted in Fig.~\ref{f2:sysdesign}. The smart meters can connect to the blockchain network by various information communication technologies, such as powerline communication, Wi-Fi, Ethernet, LoRa, and 5G Narrowband IoT. The connected blockchain nodes form a peer-to-peer network to transmit messages including the blockchain transactions via the gossip protocol.

We adopt the blockchain in the transactive energy management for three purposes. First, based on the blockchain we implement an open, verifiable, decentralized platform for the users to conduct vertical and horizontal energy transactions. Unlike the conventional centralized energy trading platform, the blockchain-based energy trading platform does not rely on a central coordinator, thus avoiding single-point failure. Second, the blockchain provides an effective and secure data communication network at a low cost. In Fig.~\ref{f2:sysdesign}, the IoT blockchain forms a peer-to-peer network than allow users to share information such as the trading decision $e_{n,m}^{\text{T}}[t]$. Third, blockchain is a convenient payment tool. The users can pay for the traded energy and FIT/DR rewards with the blockchain's build-in token.

Although there exist plenty of blockchain projects in the market, most of them were designed for PC applications. For example, running a Bitcoin full-nodes requires at least 200GB disk space, 2GB memory, 200Kbps network bandwidth, and a CPU that can support a recent version of the operating system \cite{bitcoinfull}. In this work, we consider running the blockchain nodes on smart meters based on the following two considerations. First, our energy management platform (including the blockchain) can be accommodated by the existing grid without adding new hardware. Second, running the blockchain nodes on smart meters guarantees that the energy trading data is correct and trusted, since the blockchain nodes can retrieve the trading data directly from smart meters.

However, IoT devices, including smart meters, cannot afford so much hardware resources due to limited size, power, and cost. A typical IoT device usually has an embedded CPU (e.g., ARM), memory less than 1GB, and network bandwidth less than 200Kbps (e.g., 27Kbps for LoRa). Therefore, the existing blockchain software cannot be directly deployed on IoT devices. To this end, we tailor the design of the block for IoT devices as follows. 

\subsubsection{The Consensus Protocol}\label{ss:consensus}
\begin{figure}[!t]
    \centering
    \includegraphics[width=8.7cm]{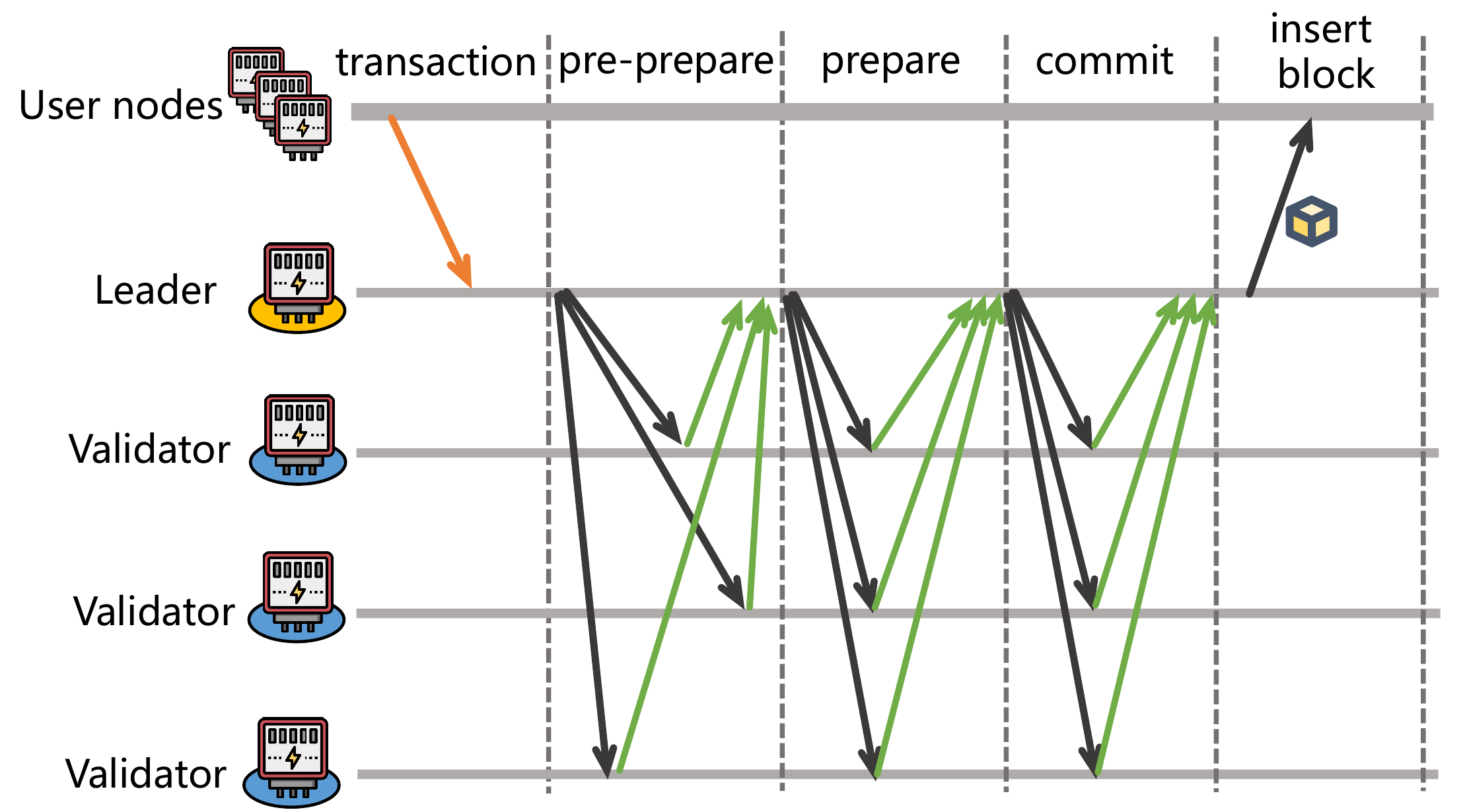}
    \caption{The modified PBFT consensus protocol for IoT blockchain.}
    \label{f3:consensus}
\end{figure}
        
\begin{figure}[!t]
    \centering
    \includegraphics[width=8.7cm]{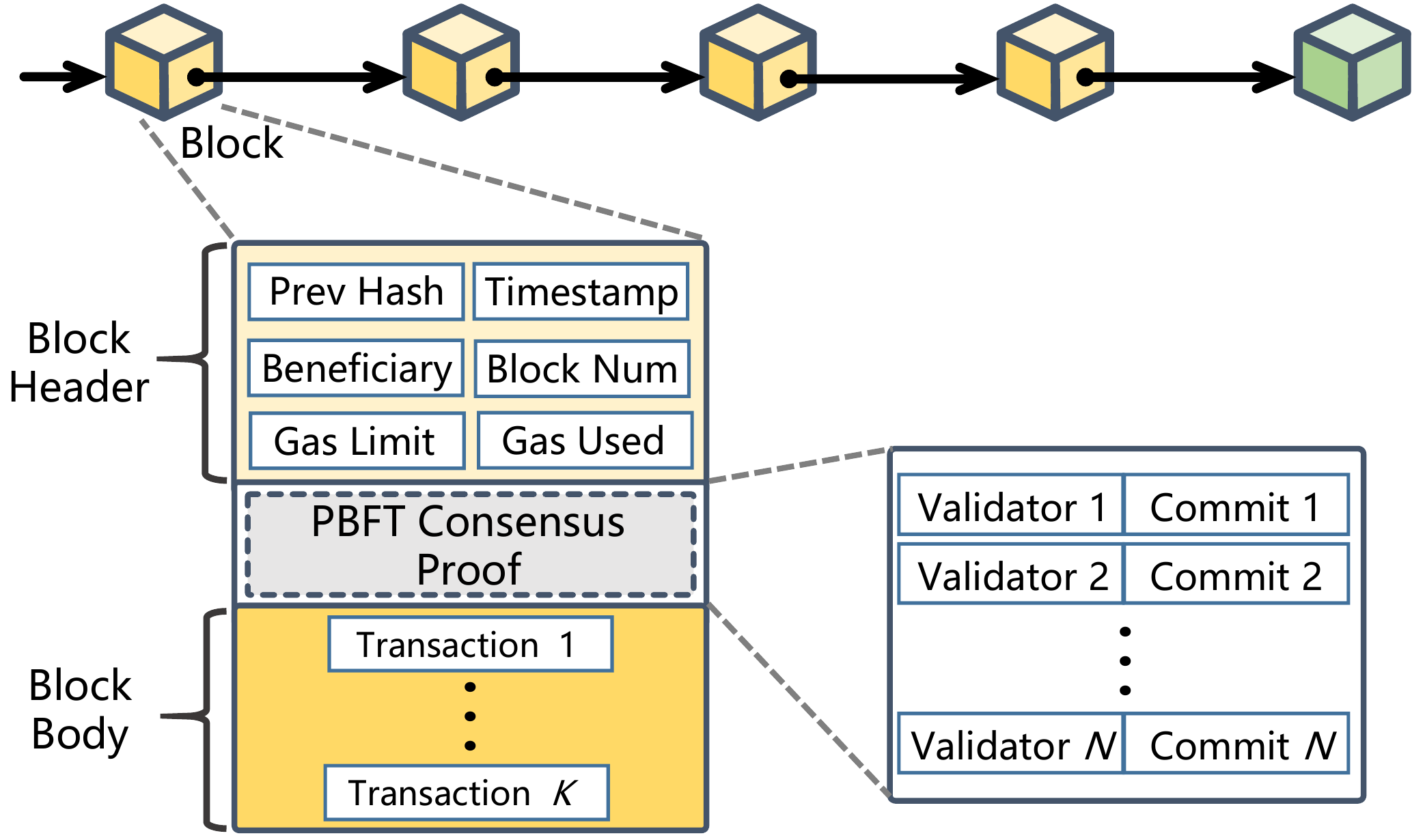}
    \caption{The block structure of the proposed IoT blockchain.}
    \label{f4:block}
\end{figure}

The blockchain node synchronizes its local state with other nodes using the consensus protocol in a distributed network. The consensus protocol is a crucial component that affects the overall performance of the blockchain system. There are many existing consensus protocols designed for different blockchains, such as PoW, PoS, DPoS, and PoA \cite{wenbo2018survey, natoli2019deconstructing}. In this work, we adopt the Byzantine fault tolerance (PBFT \cite{castro1999practical}) over other consensus protocols for the IoT blockchain based on the following three considerations. First, the computational complexity of PBFT consensus protocol is low, which is feasible on IoT devices. Second, the PBFT provides immediate finality for the transactions, which is critical for most IoT applications that require short transaction confirmation time. Third, PBFT is designed to work in asynchronous networks, so it is more resilient to the message delay and network failure, which is commonly seen in IoT networks.

To adapt the hardware of IoT devices, we develop a modified PBFT consensus protocol based on the original PBFT. The main improvements of the modified PBFT over the classical PBFT are the leader selection algorithm and the message aggregation mechanism. These improvements further reduce the complexity and increase the robustness of the consensus protocol and thus make the blockchain suitable for IoT devices.

First, we use a round-robin leader selection algorithm to choose the PBFT leader among validators. In the IoT blockchain, the nodes are classified into two types: 1) validators that participate in the consensus process to verify and generate new blocks; and 2) normal users that can emit transactions but do not participate in the consensus process. Among the validators, one is selected as the leader to lead the consensus process. As shown in Fig.\ref{f3:consensus}, the leader collects and verifies transactions from the network, initiates the three-phase communication, and generates a new block if the consensus is achieved. In conventional PBFT protocol, the leader is fixed until it fails to reach consensus. This method has a risk of single-point failure and overburdens the leader node. To address these issues, we let the validators take turns to be the leader in a pre-scheduled round-robin manner. Specifically, once the current leader successfully generates a new block, the next validator automatically becomes the new leader in the next round of consensus. This method avoids the risk of single-point failure and improves the security of the consensus process; moreover, the round-robin leader selection balances the working load of consensus among all validators, thus improving the overall efficiency of the consensus protocol.

Second, we aggregate the messages in the prepare and commit phases to reduce the consensus protocol's communication complexity. In the original PBFT protocol, validators, including the leader, must broadcast the confirmation messages with their signatures to all the other validators during the prepare and commit phases. Therefore the original protocol has a communication complexity of $O(n^2)$, which consumes high network bandwidth and prolongs the block confirmation time. In the modified PBFT protocol, we let the leader collect the confirmation message from other validators and aggregate them into a single confirmation message. Then, only the leader needs to broadcast this aggregated confirmation message to other validators during the prepare and commit phases, as shown in Fig.\ref{f3:consensus}. This method reduces the communication complexity to $O(n)$, thus saving the network bandwidth and speeding up the consensus process. We also modify the consensus proof in the block body to support the message aggregation, as shown in Fig.\ref{f4:block}.

\subsubsection{The Transaction and Block Structure}
As shown in Fig.\ref{f4:block}, we adopt the chain structure, and we choose the block structure similar to Ethereum \cite{eth}. To support the modified PBFT consensus protocol, the block contains a segment of the PBFT consensus proof. This proof is an aggregation of the commit message from all the validators so that any node can verify the block upon receiving. The block body contains all the transactions collected in the blockchain network during the consensus process.

To support the distributed transactive energy management, the blockchain has three types of transactions. The first type is the vertical energy trading transaction that carries the FIT information $e_n^{\text{FIT}}[t]$ and DR information  $e_n^{\text{DR}}[t]$. This type of transaction is made between the user and the aggregator during the FIT and DR process. The second type is the horizontal energy trading transaction that carries the trading information $e_{n,m}^{\text{T}}[t]$. This type of transaction is made by users to interact with the distributed energy trading algorithm. The third type is the token transfer transaction, which is used by the users or the aggregator to pay the rewards $R_n^{\text{FIT}}$ and $R_n^{\text{DR}}$. 

We implement the distributed transactive management with a vertical trading smart contract and a horizontal trading smart contract. The aggregator deploys the vertical trading smart contract to publish FIT and DR signals. Users interact with this smart contract to respond to the FIT and DR signals. The horizontal trading smart contract implements the transactive management algorithm. Users can call this smart contract to update their trading decisions $e_{n,m}^{\text{T}}[t]$ during the algorithm iteration, and obtain the optimal trading schedule when the algorithm converges after several rounds of iteration.

%% file: formulation.tex
Following the presentation of the system models, in this section, we formulate an optimization problem for transactive energy management. Also, we introduce three benchmark scenarios to be compared with our developed transactive energy management.

\subsection{Modeling and Optimization of the Transactive Energy Management }\label{subsec:optimization}
In the transactive-energy-management scenario, users manage their energy supply and demand locally and interact with the utility and other users. For example, as we presented in Section \ref{subsec:transactive}, users can provide extra renewable energy back to the grid and respond to demand response signals to earn rewards. Also, users can trade energy with peer users when the local renewable generation is not adequate to serve the load. By performing transactive energy management over the blockchain-enabled platform, users can fully utilize the physical and cyber connectivity and the diversity in their energy consumption behavior to benefit each other.

To better denote different costs of rewards of users, we define internal operation cost of users, rewards from vertical transaction (including feed-in renewable energy and demand response), and rewards from horizontal transaction (i.e., energy trading). To make notations concise, We denote vector forms for all the energy variables. Specifically, for the smart home loads, we define $\bm{l}_n^{\text{A}} {=} \{ l_n^{\text{A}}[t]\}$, $\bm{l}_n^{\mathrm{S}} {=} \{ l_n^{\mathrm{S}}[t]\}$, and $\bm{l}_n^{\mathrm{C}} {=} \{ l_n^{\mathrm{C}}[t]\}$; for the power supplies, we define the gird supply $\bm{s}_n^{\mathrm{G}} {=} \{ s_n^{\mathrm{G}}[t]\}$ and the renewable generation $\bm{s}_n^{\mathrm{R}} {=} \{ s_n^{\mathrm{R}}[t]\}$; for the EV charging and discharging, we define $\bm{p}_n^{\mathrm{cha}} {=} \{ p_n^{\mathrm{cha}}[t]\}$,
$\bm{p}_n^{\mathrm{dis}} {=} \{ p_n^{\mathrm{dis}}[t]\}$, and $\bm{e}_n^{\mathrm{V}} {=} \{ e_n^{\mathrm{V}}[t]\}$; for the energy trading, we define $\bm{e}_n^{\mathrm{FIT}} {=} \{ e_n^{\mathrm{FIT}}[t]\}$, $\bm{e}_n^{\mathrm{DR}} {=} \{ e_n^{\mathrm{DR}}[t]\}$, and $\bm{e}_n^{\mathrm{T}} {=} \{ e_{n,m}^{\mathrm{T}}[t]\}$. Note here the range of $t$ includes all available values within $\mathcal{T}$.

Following our analysis in Section \ref{sec:model}, the total cost incurred by user $n$ in its smart home is
\begin{align}
  \begin{aligned}
    C_n^{\text{H}}(\bm{l}_n^{\text{A}},\bm{l}_n^{\mathrm{S}},\bm{l}_n^{\mathrm{C}}, \bm{s}_n^{\mathrm{G}},\bm{p}_n^{\mathrm{dis}}) = C_n^{\text{A}}(\bm{l}_n^{\text{A}}) & + C_n^{\text{S}}(\bm{l}_n^{\mathrm{S}}) + C_n^{\text{C}}(\bm{l}_n^{\mathrm{C}}) \\
    &+ C_n^{\text{G}}(\bm{s}_n^{\mathrm{G}}) + C_n^{\text{V}}(\bm{p}_n^{\mathrm{dis}}).\label{cost-o}
  \end{aligned}
\end{align}

By participating the vertical and horizontal energy transactions, user $n$ can earn rewards to compensate the its cost. Specifically, the reward that user $n$ earns from the vertical transaction is
\begin{align}
  \begin{aligned}
    R_n^{\text{VT}}(\bm{e}_n^{\mathrm{FIT}},\bm{e}_n^{\mathrm{DR}})  = R_n^{\text{FIT}}(\bm{e}_n^{\mathrm{FIT}}) + R_n^{\text{DR}}(\bm{e}_n^{\mathrm{DR}}). \label{cost-vt}
  \end{aligned}
\end{align}
The rewards of user $n$ from horizontal transaction (i.e., energy trading) is shown in \eqref{objective-et}, i.e., $R_n^{\text{T}}(\bm{e}_n^{\mathrm{T}})$. By separately denoting the operation cost and rewards, we can consider different scenarios, where users only optimize the internal operations and jointly optimize both internal operations and external transactions. We will focus on the full transaction in this section and present other benchmark scenarios in Section \ref{sec:benchmark}. 

Since the total supply and demand must be balanced for the smart home, the transactive energy management of user $n$ should always satisfy the 
the energy balance constraint below
    \begin{equation}
        \begin{aligned}
            l_n^{\text{A}}[t] + l_n^{\text{S}}[t] + & l_n^{\text{C}}[t] + l_n^{\text{I}}[t] + p_n^{\text{cha}}[t] + \sum\nolimits_{m \in \mathcal{N} \backslash n} e_{n,m}^{\text{T}}[t] \\
            & = s_n^{\text{R}}[t] + s_n^{\text{G}}[t] - e_n^{\text{DR}}[t] + p_n^{\text{dis}}[t] , ~ \forall t \in \mathcal{T}, \label{constraint-load12}
        \end{aligned}
    \end{equation}
where the left-hand side is the total demand of user $n$ including all types of loads, EV charging, and energy sold to others $\sum_{m \in \mathcal{N} \backslash n} e_{n,m}^{\text{T}}[t]$. The right-hand side is the total supply adjusted by the committed demand response $e_n^{\text{DR}}[t]$.

All the users seek to minimize their costs by optimizing their load schedule, renewable generation supply, EV charging, and participation in energy transactions.
We consider the optimization of transactive energy management from the perspective of the system and aim to minimize the total costs of all the users. Therefore, we define the overall cost optimization target for the IoT blockchain-based transactive energy management system as

\noindent\textbf{TEM}: Transactive Energy Management
    \begin{equation*}
        \begin{aligned}
            &\text{minimize} {}&&{} \sum_{n\in\mathcal{N}} C_n^{\text{H}}(\bm{l}_n^{\text{A}},\bm{l}_n^{\mathrm{S}},\bm{l}_n^{\mathrm{C}},\bm{s}_n^{\mathrm{G}},\bm{p}_n^{\mathrm{dis}}) -  \sum_{n\in\mathcal{N}} R_n^{\text{T}}(\bm{e}_n^{\mathrm{T}})\\
            &&& \quad\quad\quad\quad\quad\quad - \sum_{n\in\mathcal{N}} R_n^{\text{VT}}(\bm{e}_n^{\mathrm{FIT}},\bm{e}_n^{\mathrm{DR}})\\
            &\text{subject to} {}&&{} 
            \text{\eqref{constraint-load3}},\text{\eqref{constraint-load5}},\text{\eqref{constraint-load1}},\text{\eqref{constraint-ev2}},\text{\eqref{constraint-ev5}}, \text{\eqref{constraint-load8}}, \eqref{constraint-load9}, \text{\eqref{constraint-load10}}, \text{\eqref{constraint-load11}}, \text{\eqref{constraint-load12}}\\            
            &\text{variables:} {}&&{}
            \{ \bm{l}_n^{\text{A}},\bm{l}_n^{\mathrm{S}},\bm{l}_n^{\mathrm{C}},\bm{s}_n^{\mathrm{G}},\bm{s}_n^{\mathrm{R}},\bm{p}_n^{\mathrm{dis}},\bm{e}_n^{\mathrm{FIT}},\bm{e}_n^{\mathrm{DR}},\bm{e}_n^{\mathrm{T}} \}.
        \end{aligned} 
    \end{equation*}

The decision variables in TEM include internal energy scheduling decisions in each smart home $\{\bm{l}_n^{\text{A}},\bm{l}_n^{\mathrm{S}},\bm{l}_n^{\mathrm{C}},\bm{s}_n^{\mathrm{G}},\bm{s}_n^{\mathrm{R}},\bm{p}_n^{\mathrm{dis}}\}$ and external transactive energy decisions $\{\bm{e}_n^{\mathrm{FIT}},\bm{e}_n^{\mathrm{DR}},\bm{e}_n^{\mathrm{T}}\}$ that interact with the grid and other users. Since the decision variables are coupled across all users, the traditional way of solving Problem TEM is to let a central coordinator collect all the users' information and solve TEM in a centralized manner. However, the centralized method raises serious privacy concerns because all users reveal the above private information to the central coordinator. To address the privacy concern, we develop a privacy-persevering optimization method to solve TEM, which will be presented in the next section.

\subsection{Distributed Optimization Method for TEM}\label{sec:algorithm}
To preserve the users' privacy while obtaining the optimal solution for TEM, we design a distributed optimization algorithm that can be implemented using the smart contract of the IoT blockchain. First, we adopt the alternating direction method of multipliers (ADMM) method \cite{boyd2011distributed} to decompose TEM into two tasks: the user local task (ULT) and the smart contract task (SCT). The ULT is run by users locally to generate users' optimized power usage schedule and transactive-energy decisions. The SCT is run on the IoT blockchain as a smart contract that collects the users' local trading decisions and leads to the globally optimal trading decision. The users' private information is only used by the ULT locally; therefore, the users do not reveal privacy to other parties. The IoT blockchain guarantees that the SCT results are correct and intact because SCT is implemented as a smart contract whose operation cannot be intervened by anybody. The design of the distributed algorithm is described as follows.
    
The ADMM method \cite{boyd2011distributed} is an algorithm is a promising distributed algorithm used in energy trading \cite{wang2016incentivizing} for its good convergence and scalability. According to the ADMM method, we first define $\hat{\bm{e}}_n^{\text{T}} {=} \{\hat{e}_{n,m}^{\text{T}}[t], \forall m {\in} \mathcal{N} \backslash n\}$ as the auxiliary variable of the horizontal transactive energy decisions $\bm{e}_n^{\text{T}}$. Based on constraint \eqref{constraint-load11}, we can obtain the equivalent constraints as
    \begin{align}
        \hat{e}_{n,m}^{\text{T}}[t] &= e_{n,m}^{\text{T}}[t], \forall m {\in} \mathcal{N} \backslash n,\forall n {\in} \mathcal{N},\forall t {\in} \mathcal{T}, \label{constraint-auxiliary1}\\
        \hat{e}_{n,m}^{\text{T}}[t] {+} \hat{e}_{m,n}^{\text{T}}[t] &= 0, \forall m {\in} \mathcal{N} \backslash n,\forall n {\in} \mathcal{N},\forall t \in \mathcal{T}. \label{constraint-auxiliary2}
    \end{align}

Next we apply the augmented Lagrangian method on TEM by defining the dual variables $\bm{\lambda}_{n} {=} \{ \lambda_{n,m}^{t},\forall m {\in} \mathcal{N} \backslash n,t {\in} \mathcal{T}\}$ for constraints \eqref{constraint-auxiliary1}. Then we define a new positive parameter $\rho$ as the weight of the penalty in \eqref{constraint-auxiliary1}. The augmented Lagrangian is
\begin{equation}
    \begin{aligned}    
        L 
        &= \sum_{n\in\mathcal{N}} C_n^{\text{H}}(\bm{l}_n^{\text{A}},\bm{l}_n^{\mathrm{S}},\bm{l}_n^{\mathrm{C}},\bm{s}_n^{\mathrm{G}},\bm{p}_n^{\mathrm{dis}}) \\
        &- \sum_{n\in\mathcal{N}} \left(  R_n^{\text{VT}}(\bm{e}_n^{\mathrm{FIT}},\bm{e}_n^{\mathrm{DR}}) + R_n^{\text{T}}(\bm{e}_n^{\mathrm{T}}) \right) \\
        &{+} \!\! {\sum_{n\in\mathcal{N}}} {\sum_{m \in \mathcal{N} \backslash n}} \! \sum_{t\in\mathcal{T}} 
        \Big[ \frac{\rho}{2} \left( \hat{e}_{n,m}^{\text{T}}[t] {-} e_{n,m}^{\text{T}}[t] \right)^{2} \\
        & {+} \lambda_{n,m}^{t} \left( \hat{e}_{n,m}^{\text{T}}[t] {-} e_{n,m}^{\text{T}}[t] \right) 
        \Big]. \label{constraint-lagrangian}
    \end{aligned}
\end{equation}

From \eqref{constraint-lagrangian}, we observe that given auxiliary variables $\hat{e}_{n,m}^{\text{T}}[t]$, each user can solve an individual optimization problem with all decisions decomposed from other users. Therefore, we decompose TEM into two tasks, namely, the user local task (ULT) and the smart contract task (SCT). ULT locally optimizes the users' costs individually and outputs the trading decisions to the SCT. The SCT is implemented in a smart contract that collects the users' trading decisions to calculate the auxiliary variables and dual variables and feed them back to the users. During this process, the users send transactions to the smart contract to communicate the values of the variables.

After user $n$ obtain the latest value of $\lambda_{n,m}^{t}$ and $\hat{e}_{n,m}^{\text{T}}[t]$, it works on the following optimization task

\noindent\textbf{ULT$_n$}: User Local Task
    \begin{equation*}
        \begin{aligned}
            & \text{minimize} 
            && \sum_{n\in\mathcal{N}} C_n^{\text{H}}(\bm{l}_n^{\text{A}},\bm{l}_n^{\mathrm{S}},\bm{l}_n^{\mathrm{C}},\bm{s}_n^{\mathrm{G}},\bm{p}_n^{\mathrm{dis}}) \\
          &&& - \sum_{n\in\mathcal{N}} \left(  R_n^{\text{VT}}(\bm{e}_n^{\mathrm{FIT}},\bm{e}_n^{\mathrm{DR}}) + R_n^{\text{T}}(\bm{e}_n^{\mathrm{T}}) \right) \\
          &&& {+} \!\! \sum_{m \in \mathcal{N} \backslash n} \! \sum_{t\in\mathcal{T}} 
        \left[ \frac{\rho}{2} \left( \hat{e}_{n,m}^{\text{T}}[t] {-} e_{n,m}^{\text{T}}[t] \right)^{2} 
        {-} \lambda_{n,m}^{t}  e_{n,m}^{\text{T}}[t] \right] \\
            &\text{subject to} &&
            \text{\eqref{constraint-load3}},\text{\eqref{constraint-load5}},\text{\eqref{constraint-load1}},\text{\eqref{constraint-ev2}},\text{\eqref{constraint-ev5}}, \eqref{constraint-load8}, \eqref{constraint-load9}, \text{\eqref{constraint-load10}},\text{\eqref{constraint-load12}}\\
            &\text{variables:} &&
            \bm{l}_n^{\text{A}}, \bm{l}_n^{\mathrm{S}},\bm{l}_n^{\mathrm{C}},\bm{s}_n^{\mathrm{G}},\bm{s}_n^{\mathrm{R}},\bm{p}_n^{\mathrm{dis}},\bm{e}_n^{\mathrm{FIT}},\bm{e}_n^{\mathrm{DR}},\bm{e}_n^{\mathrm{T}}.
        \end{aligned} 
    \end{equation*}
    
User $n$ solve ULT$_n$ to obtain its locally optimal energy management including the energy trading decision $\bm{e}_n^{\text{T}}$. Then the user calls the smart contract of SCT to update its trading decision for the next iteration.

Similarly, the smart contract, upon receiving all the users' trading decisions, works on the following optimization task    

\noindent\textbf{SCT}: Smart Contract Task
    \begin{equation*}
        \begin{aligned}
        &\text{minimize} && \sum_{n\in\mathcal{N}} \sum_{m \in \mathcal{N} \backslash n} \sum_{t\in\mathcal{T}} 
        \Big\{ \frac{\rho}{2} \left( \hat{e}_{n,m}^{\text{T}}[t] - e_{n,m}^{\text{T}}[t] \right)^{2} \\
        &&&  \quad\quad\quad\quad + \lambda_{n,m}^{t} \hat{e}_{n,m}^{\text{T}}[t] \Big\} \\
        &\text{subject to} &&
        \text{\eqref{constraint-auxiliary2}} \\
        &\text{variables:} &&
        \{ \hat{\bm{e}}_n^{\text{T}},~ \forall n \in \mathcal{N} \}.
        \end{aligned} 
    \end{equation*}

The SCT updates the dual variables $\boldsymbol{\lambda} {=} \{ \bm{\lambda}_{n},\forall n {\in} \mathcal{N}\}$ and auxiliary variables $\boldsymbol{\hat{e}}_n^{\text{T}}$, and the users can obtain the lasted value of $\boldsymbol{\lambda}$ and $\boldsymbol{\hat{e}}_n^{\text{T}}$ by accessing the smart contract. Specifically, SCT calculates the optimal auxiliary variables by
        \begin{align}
            \begin{split}
                \hat{e}_{n,m}^{\text{T}}[t] {=} -\hat{e}_{m,n}^{\text{T}}[t]
                {=} \frac{\rho \left( e_{n,m}^{\text{T}}[t] - e_{m,n}^{\text{T}}[t] \right) - \left( \lambda_{n,m}^{t} - \lambda_{m,n}^{t} \right) }{2 \rho}, \label{updateenergy}
            \end{split}
        \end{align}
and updates the dual variables as
    \begin{align}
        \lambda_{n,m}^{t} \leftarrow \lambda_{n,m}^{t} + \rho \left( \hat{e}_{n,m}^{\text{T}}[t] - e_{n,m}^{\text{T}}[t] \right). \label{updatelambda}
    \end{align}

We implement SCT in the smart contract that is deployed on the blockchain. The smart contract, which is implemented in Solidity, consists of three core functions. The first function is to solve the optimization problem of SCT by implementing the numerical computation of \eqref{updateenergy} and \eqref{updatelambda}. The second function is to set new values to the variables of the users' energy trading decisions $\bm{e}_n^{\text{T}}$. The users can call this function to update their local trading decisions in each iteration of Algorithm~\ref{alg1}. The third function is to reveal the values of the dual variables $\boldsymbol{\lambda}$ and auxiliary variables $\boldsymbol{\hat{e}}_n^{\text{T}}$. The users can call this function to read the latest values of $\boldsymbol{\lambda}$ and $\boldsymbol{\hat{e}}_n^{\text{T}}$ in each iteration.

By decomposing TEM into ULT and SCT, we obtain a distributed solution to the optimization problem of TEM. More importantly, the ULT can be locally solved by the users in a parallel manner; and the SCT is implemented by the smart contract and guaranteed to be accurate and tamper-proof. The blockchain provides a reliable communication network and a trusted computing machine to solve the optimization problem of TEM. First, the information exchange between ULT$_n$ and SCT of Algorithm~\ref{alg1} is conducted over the blockchain. Second, the SCT part of Algorithm~\ref{alg1} is implemented in the smart contract on the blockchain. The blockchain acts as a trusted computing machine that solves the optimization problem of SCT, and thus removes the need for a central coordinator.

The proposed energy management algorithm preserves the users' privacy by minimizing the amount of information that the user needs to reveal to other parties. As shown in Algorithm~\ref{alg1}, the distributed transactive energy management algorithm works iteratively. During the iteration of the distributed algorithm, the users do not need to reveal the process of optimizing the trading decisions, so that their privacy is well preserved. Moreover, Algorithm~\ref{alg1} is guaranteed to converge to the optimal solution because the original optimization target of TEM is convex. To guarantee the convergence of Algorithm~\ref{alg1}, $\rho(k)$ is chosen to be the reciprocal of the number of iteration.

\begin{algorithm}[!t]
     \caption{The distributed transactive energy management algorithm}
     \label{alg1}
     \SetAlgoLined
     \textbf{Initialization}:\\
     \hspace*{1em} iteration number $k {\leftarrow} 1$; step size $\rho(0) {\leftarrow} 1$; \\
     \hspace*{1em} auxiliary variable $\boldsymbol{\lambda}(0) {\leftarrow} \textbf{0}$; \\
     \hspace*{1em} convergence error thresholds ${\epsilon \leftarrow} 1\times10^{-6}$; \\
     \hspace*{1em} deploy the smart contract of SCT;
    
    \While{ $\sum_{n \in \mathcal{N}} \parallel \bm{\hat{e}}_n^{\mathrm{T}}(k) - \bm{e}_n^{\mathrm{T}}(k) \parallel > \epsilon$ or $\parallel \bm{\lambda}(k) - \bm{\lambda}(k-1) \parallel > \epsilon$ }{
    \hspace*{-1.1em}$\neg$ \For{$n \in \mathcal{N}$}{
        \hspace*{-1.1em}$\neg$ User $n$ access SCT to obtain $\bm{\hat{e}}_{n}^{\text{T}}(k)$ and $\bm{\lambda}(k)$;    
    
        \hspace*{-1.1em}$\neg$ User $n$ solves task ULT$_n$ based on $\hat{\bm{e}}_n^{\text{T}}(k{-}1)$ and $\bm{\lambda}_n(k{-}1)$;
        
        \hspace*{-1.1em}$\neg$ User $n$ updates $\bm{e}_n^{\text{T}}(k)$ to SCT;
    }
    
    \hspace*{-1.1em}$\neg$ The smart contract executes to solve SCT 
    
    \hspace*{-1.1em}$\neg$ The smart contract updates $\bm{\hat{e}}_{n}^{\text{T}}(k),\forall n \in \mathcal{N}$ and $\bm{\lambda}(k)$;
    
    \hspace*{-1.1em}$\neg$ $k \leftarrow k+1$;
    }
 \KwResult{the optimal energy trading schedule $\bm{e}_n^{\text{T},\ast},~\forall n {\in} \mathcal{N}$.}
\end{algorithm}

\subsection{Benchmark Scenarios}\label{sec:benchmark}
After we formulate the optimization problem TEM for the transactive energy management in Section~\ref{subsec:optimization} and solve it in a distributed manner in Section~\ref{sec:algorithm}, we are interested in comparing TEM with the following benchmark scenarios.
\begin{enumerate}
\item \textit{Benchmark Scenario 1}: Users do not participate in any transactive energy activities and only optimize their internal energy schedule alone.
\item \textit{Benchmark Scenario 2}: Users only participate in vertical transactions, including feed-in renewable and demand response, and jointly optimize the internal energy schedule and vertical transactions.
\item \textit{Benchmark Scenario 3}: Users only participate in horizontal transactions with other users and jointly optimize the internal energy schedule and horizontal transactions.
\end{enumerate}

We can show a comprehensive evaluation of different transactive energy schemes and the corresponding benefits to users and the whole system through comparisons with the above three benchmark scenarios. 

Specifically, in benchmark scenario 1, user $n$ only schedules energy supply and demand in the smart home, and thus user $n$ needs to balance the total power supply and demand as follows.
    \begin{equation}
    \begin{aligned}
            & l_n^{\text{A}}[t] + l_n^{\text{S}}[t] + l_n^{\text{C}}[t] + l_n^{\text{I}}[t] + p_n^{\text{cha}}[t] \\
            & = s_n^{\text{R}}[t] + s_n^{\text{G}}[t] + p_n^{\text{dis}}[t] , ~ \forall n \in \mathcal{N}, t \in \mathcal{T}. \label{constraint-load13}
    \end{aligned}
    \end{equation}
    
We can formulate the energy management problem for benchmark scenario 1 as

\textbf{BS1}: Optimization Problem for Benchmark Scenario 1
    \begin{equation*}
        \begin{aligned}
            &\text{minimize} {}&&{} \sum_{n\in\mathcal{N}} C_n^{\text{H}}(\bm{l}_n^{\text{A}},\bm{l}_n^{\mathrm{S}},\bm{l}_n^{\mathrm{C}},\bm{s}_n^{\mathrm{G}},\bm{p}_n^{\mathrm{dis}})  \\ &\text{subject to} {}&&{} 
             \text{\eqref{constraint-load3}},\text{\eqref{constraint-load5}}, \text{\eqref{constraint-load1}}, \text{\eqref{constraint-ev2}},\text{\eqref{constraint-ev5}},\text{\eqref{constraint-load13}}\\
            &\text{variables:} {}&&{}
            \{ \bm{l}_n^{\text{A}},\bm{l}_n^{\mathrm{S}},\bm{l}_n^{\mathrm{C}},\bm{s}_n^{\mathrm{G}},\bm{s}_n^{\mathrm{R}},\bm{p}_n^{\mathrm{dis}},~\forall n \in \mathcal{N} \}.
        \end{aligned} 
    \end{equation*}

Similarly, for benchmark scenario 2, we present the energy balance constraints as 
    \begin{equation}
        \begin{aligned}
            & l_n^{\text{A}} + l_n^{\text{S}}[t] + l_n^{\text{C}}[t] + l_n^{\text{I}}[t] + p_n^{\text{cha}}[t] \\
            & = s_n^{\text{R}}[t] + s_n^{\text{G}}[t] - e_n^{\text{DR}}[t] + p_n^{\text{dis}}[t] , ~ \forall n \in \mathcal{N}, t \in \mathcal{T}, \label{constraint-load14}
        \end{aligned}
    \end{equation}
and formulate the energy management problem for benchmark scenario 2 as

\textbf{BS2}: Optimization Problem for Benchmark Scenario 2
\begin{equation*}
        \begin{aligned}
            &\text{minimize} {}&&{} \sum_{n\in\mathcal{N}} C_n^{\text{H}}(\bm{l}_n^{\text{A}},\bm{l}_n^{\mathrm{S}},\bm{l}_n^{\mathrm{C}},\bm{s}_n^{\mathrm{G}},\bm{p}_n^{\mathrm{dis}}) \\
            &&& - \sum_{n\in\mathcal{N}}  R_n^{\text{VT}}(\bm{e}_n^{\mathrm{FIT}},\bm{e}_n^{\mathrm{DR}}) \\
            &\text{subject to} {}&&{} 
            \text{\eqref{constraint-load3}},
            \text{\eqref{constraint-load5}}, \text{\eqref{constraint-load1}},
            \text{\eqref{constraint-ev2}},\text{\eqref{constraint-ev5}}, \text{\eqref{constraint-load8}-\eqref{constraint-load9}}, \text{\eqref{constraint-load10}}, \text{\eqref{constraint-load14}}\\
            &\text{variables:} {}&&{}
            \{ \bm{l}_n^{\text{A}},\bm{l}_n^{\mathrm{S}},\bm{l}_n^{\mathrm{C}},\bm{s}_n^{\mathrm{G}},\bm{s}_n^{\mathrm{R}},\bm{p}_n^{\mathrm{dis}},\bm{e}_n^{\mathrm{FIT}},\bm{e}_n^{\mathrm{DR}},~\forall n \in \mathcal{N} \}.
        \end{aligned} 
\end{equation*}

For benchmark scenario 3, we have the following energy balance constraints
    \begin{equation}
        \begin{aligned}
            & l_n^{\text{A}}[t] + l_n^{\text{S}}[t] + l_n^{\text{C}}[t] + l_n^{\text{I}}[t] + p_n^{\text{cha}}[t] + \sum\nolimits_{m \in \mathcal{N} \backslash n} e_{n,m}^{\text{T}}[t] \\
            & = s_n^{\text{R}}[t] + s_n^{\text{G}}[t] + p_n^{\text{dis}}[t] , ~ \forall n \in \mathcal{N}, t \in \mathcal{T}, \label{constraint-load15}
        \end{aligned}
    \end{equation}
and formulate the energy management problem as

\textbf{BS3}: Optimization Problem for Benchmark Scenario 3
\begin{equation*}
        \begin{aligned}
            &\text{minimize} {}&&{} \sum_{n\in\mathcal{N}} \left(
            C_n^{\text{H}}(\bm{l}_n^{\text{A}},\bm{l}_n^{\mathrm{S}},\bm{l}_n^{\mathrm{C}},\bm{s}_n^{\mathrm{G}},\bm{p}_n^{\mathrm{dis}}) - R_n^{\text{T}}(\bm{e}_n^{\mathrm{T}}) \right)\\
            &\text{subject to} {}&&{} 
            \text{\eqref{constraint-load3}},
            \text{\eqref{constraint-load5}},
            \text{\eqref{constraint-load1}},
            \text{\eqref{constraint-ev2}},\text{\eqref{constraint-ev5}}, \text{\eqref{constraint-load11}}, \text{\eqref{constraint-load15}}\\
            &\text{variables:} {}&&{}
            \{ \bm{l}_n^{\text{A}},\bm{l}_n^{\mathrm{S}},\bm{l}_n^{\mathrm{C}},\bm{s}_n^{\mathrm{G}},\bm{s}_n^{\mathrm{R}},\bm{p}_n^{\mathrm{dis}},\bm{e}_n^{\mathrm{T}},~\forall n \in \mathcal{N} \}.
        \end{aligned} 
\end{equation*}

Since users do not participate in horizontal transactions in BS1 and BS2, they are naturally decoupled with each other. Therefore, user $n$ can solve its energy management problem individually using standard convex optimization techniques. In BS3, users' trading decisions are coupled so that we can follow the similar steps in Section \ref{sec:algorithm} to solve BS3 using the distributed algorithm. Since the benchmark scenarios are designed for comparisons with our transactive energy management algorithm for TEM, we skip the solution method for three benchmark problems due to the page limit.

%% file: evaluation.tex
This section consists of two parts: 1) a systematic test of the proposed blockchain design for transactive energy management on a realistic network of IoT devices in Fig.~\ref{f5:demo}; and 2) evaluation and analysis of our blockchain-based transactive energy management algorithm by conducting numerical simulations with data collected from practical applications.

\subsection{Performance Evaluation of the IoT Blockchain} \label{sec:blockchain}

\subsubsection{Experiment Setup} 
\begin{figure}[!t]
    \centering
    \includegraphics[width=8.7cm]{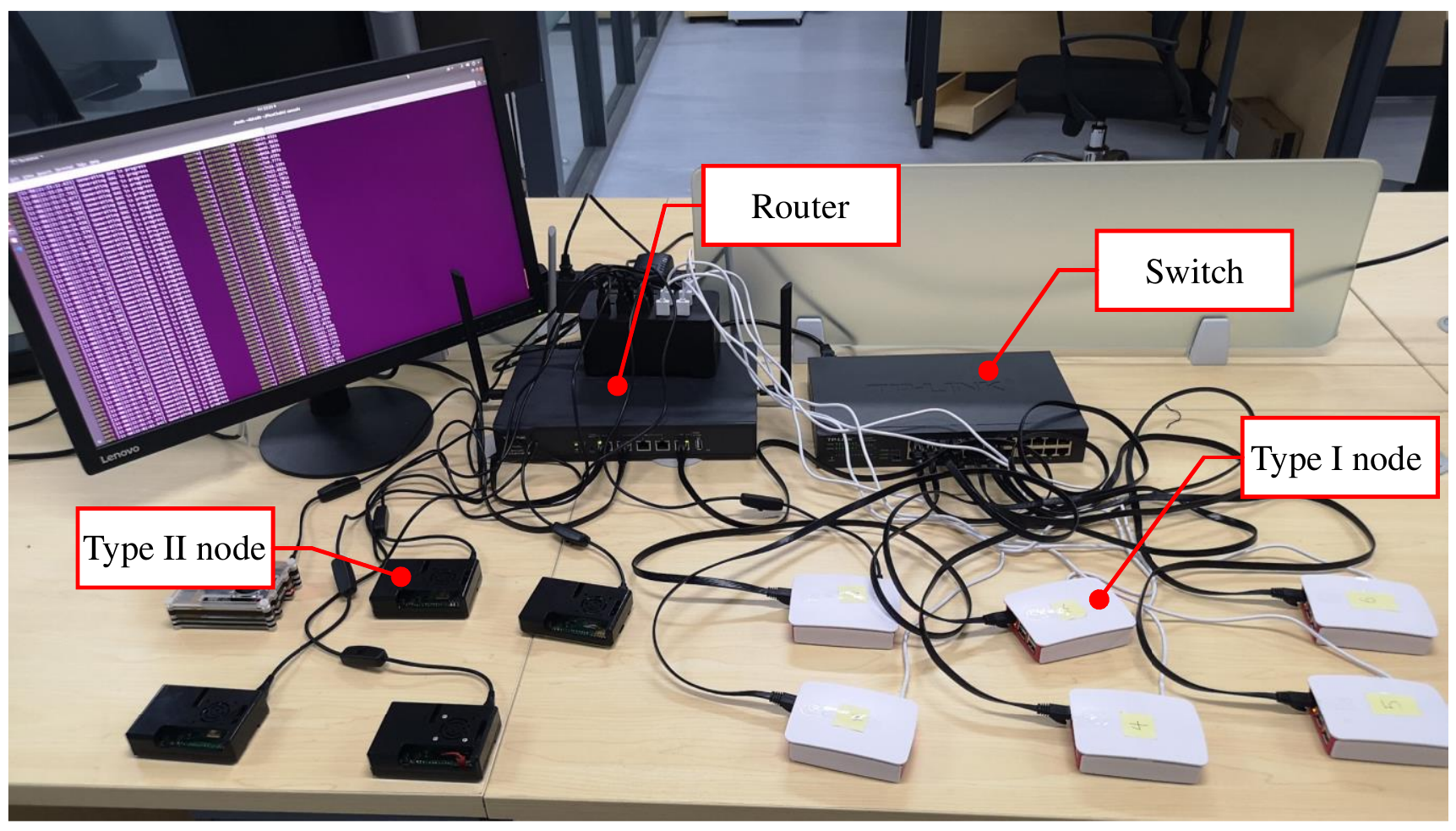}
    \caption{The test network of IoT devices for the evaluation of the blockchain system.}
    \label{f5:demo}
\end{figure}

We build a test network of 11 Raspberry Pis to evaluate the IoT blockchain designed in Section~\ref{sec:blockchainlayer}. As shown in Fig.~\ref{f5:demo}, we use two types of Raspberry Pi \cite{rpi} to emulate the high-end and low-end IoT devices. Specifically, the type-I node is a Raspberry~Pi Model~3B+ module with a Broadcom BCM2837B0 CPU (quadcore A53 at 1.4GHz) and 1GB DDR2 SDRAM; type-II node is a Raspberry~Pi Model~2B module with Broadcom BCM2836 CPU (quadcore A7 at 900MHz) and 512MB DDR2 SDRAM. We use a switch and a router (both from TPlink) to connect the Raspberry~Pis to form a local private network. The router is used to limit the network bandwidth of the Raspberry~Pis to be less than 250Kbps. 

\subsubsection{The IoT Blockchain Evaluation}
We implement the IoT blockchain design based on the source code of Quorum \cite{quorum}. Quorum is a modified version of Ethereum for FinTech applications. Quorum modified the PoW consensus protocol of Ethereum to the PBFT protocol, but other parts remain the same as Ethereum. We further modify Quorum to implement the message aggregation described in Section~\ref{sec:blockchainlayer}. The final binary file of the IoT blockchain is 36MB using the Go language compiler. Since Quorum supports smart contracts, we use Solidity to implement the SCT part of our proposed distributed algorithm for TEM.

We compare the resource consumption of the proposed IoT blockchain with Ethereum. For the IoT blockchain, running a full-node validator node consumes 480MB memory, and running a normal node consumes about 200MB memory. By contrast, for Ethereum, a full node consumes 400MB memory without mining and 1GB memory with mining. On the Raspberry~Pi Model~3B+ module, the CPU consumption of IoT blockchain validator is less than 50\%; however, Ethereum mining consumes 100\% of the CPU time, which makes the operating system very slow. Our test shows that the IoT blockchain can run smoothly on the IoT devices.

To test the throughput of the IoT blockchain, we let five type-I nodes to be the validators and five type-II nodes to be the normal users. We use one type-II node to monitor the transactions and blocks in the blockchain. Our test shows that the delay of the transaction is about 5ms, and the block confirmation time is less than 100ms. The measured highest TPS (transaction per second) is around 700 in the network. The results show that the IoT blockchain's performance is sufficient to support the execution of the distributed transactive energy management algorithm. To evaluate the SCT algorithm, we run Algorithm~\ref{alg1} in Matlab and update the results to the smart contract in each iteration. 

\subsection{Numerical Simulations of the Distributed Transactive Energy Management Algorithm}
\subsubsection{Simulation Data and Parameters}
The energy data used in our simulations comes from \cite{wang2015joint} and \cite{pecan}, including power consumption, renewable energy generation (e.g., solar and wind), and outdoor temperature from September 6 to September 12 in 2016. The transactive energy algorithm is executed to determine the day-ahead energy scheduling, and the payment is settled at the end of the trading day.\footnote{The setup for the transactive energy is aligned with the practice of the day-ahead market in power grids. Market participants bid their trading decisions based on their prediction of the market parameters and their operational parameters, e.g., generations and loads, one day before the actual trading day.} The simulation parameters are listed in Table~\ref{t1:parameter}.

\begin{table}[!t]
    \centering
    \renewcommand{\arraystretch}{1.3}
    \caption{Parameters used in the numerical simulation}
    \label{t1:parameter}
    \resizebox{\columnwidth}{!}{
    \begin{tabular}{l l l}
        \hline\hline \\[-3mm]
        \multicolumn{1}{c}{Parameter} & \multicolumn{1}{c}{Value} & \multicolumn{1}{c}{\pbox{20cm}{Description}}  \\[1.2ex] \hline

     $\alpha$, $\beta$   & 0.75, 0.2  & {Working efficiency of the HVAC system}\\
     $\omega_{\text{A}}$ & 1 & {User's sensitivity to HAVC}\\
     $\text{Tout}^n[t]$ & From history data & {Outdoor temperature}\\
     $\underline{\text{Tin}}^n$ and $\overline{\text{Tin}}^n$ & $15^\circ$C, $32^\circ$C, & {The upper/lower bound of indoor temperature}\\
     $L_n^{\text{S}}[t]$ & From dataset \cite{pecan} & {The shiftable load preference of user $n$ at time $t$}\\
     $\omega_{\text{S}}$  & 1 & {Users' sensitivity of the behavior change due to shifted load}\\
     $S_n^{\text{R}}[t]$ & From real data & {The upper bound of renewable energy generation of user $n$ at time $t$}\\
     $S_{\text{G}}$  & 20kWh & {The upper bound of power draw from the grid powerline at time $t$}\\
     $p_{\text{G}}$,  $p^{\star}_{\text{G}}$ & 0.2, 0.8 & {The normal and peak price of the grid}\\
     $\mu_n$, $\nu_n$ & 0.9, 0.9 & {The charging/discharging efficiency of the EV}\\
     $E_n^{\text{V}}$ & Random in [30kWh, 50kWh] & {The battery capacity of user $n$'s EV}\\
     $P_n^{\text{cha}}$ and $P_n^{\text{dis}}$ & 50kWh, 10kWh & {Upper bound of the charging/discharging power per hour of user $n$'s EV}\\
     $\omega_{\text{V}}$ & 0.1 & {The cost coefficient of the EV battery cost}\\ 
     $[t_n^{\text{A}},t_n^{\text{D}}]$ & [9, 18] & {The departure time and arrival time of user $n$'s EV}\\ [1.2ex]
    \hline\hline
    \end{tabular}
    }
\end{table}

\begin{figure}[!t]
    \centering
    \includegraphics[width=8.5cm]{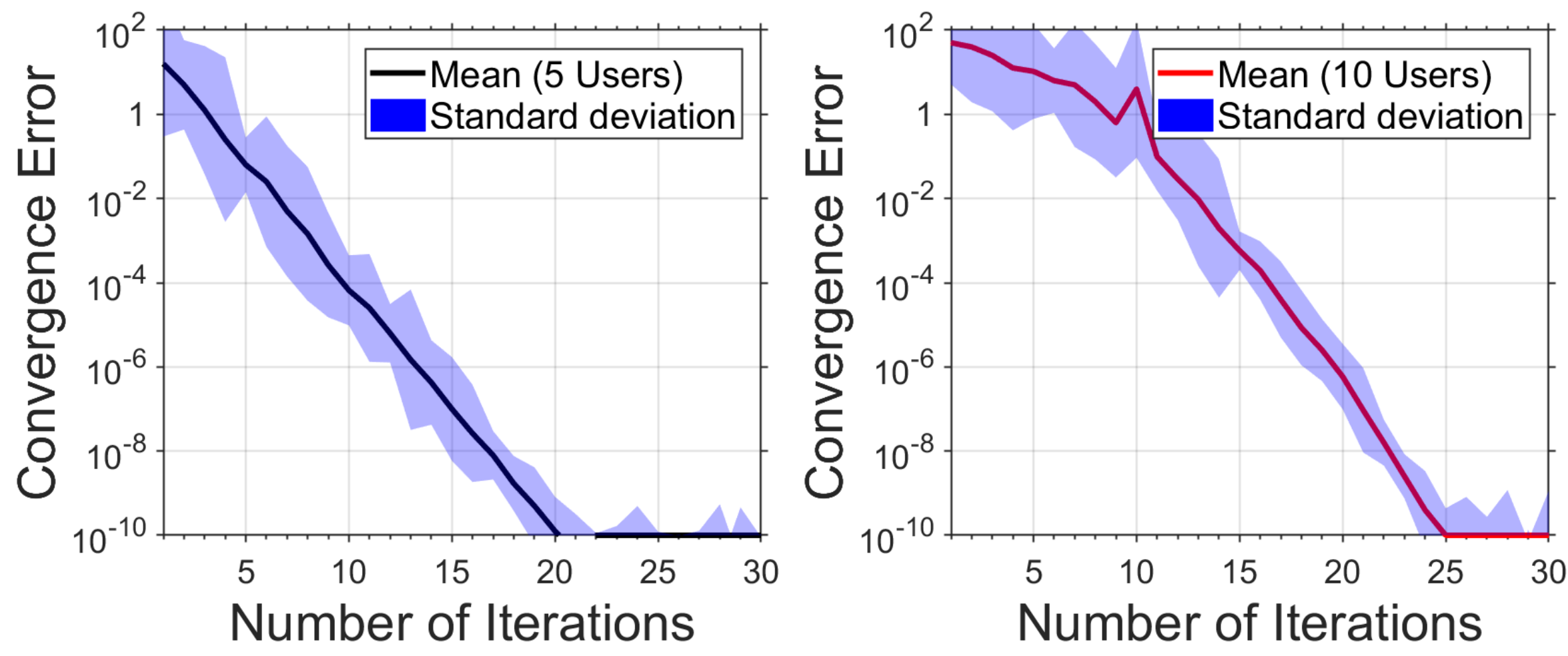}
    \caption{The convergence error of the distributed transactive energy management algorithm. We plot cases with 5 users (left) and 10 users (right).}
    \label{f:converge}
\end{figure}

\subsubsection{Algorithm Convergence}
Since the distributed transactive energy management algorithm consists of two tasks, it works in an iterative manner. To show its convergence performance, we simulate the algorithm in two cases with five users and ten users. We set the threshold of the convergence error $\epsilon = 1\times10^{-10}$ in Algorithm~\ref{alg1}. We plot the convergence in both cases in Fig.~\ref{f:converge}. The results show that Algorithm~\ref{alg1} converges at 16th iteration for 5 users, and at 22nd iteration for 10 users.

\begin{figure}[!t]
    \centering
    \includegraphics[width=8.7cm]{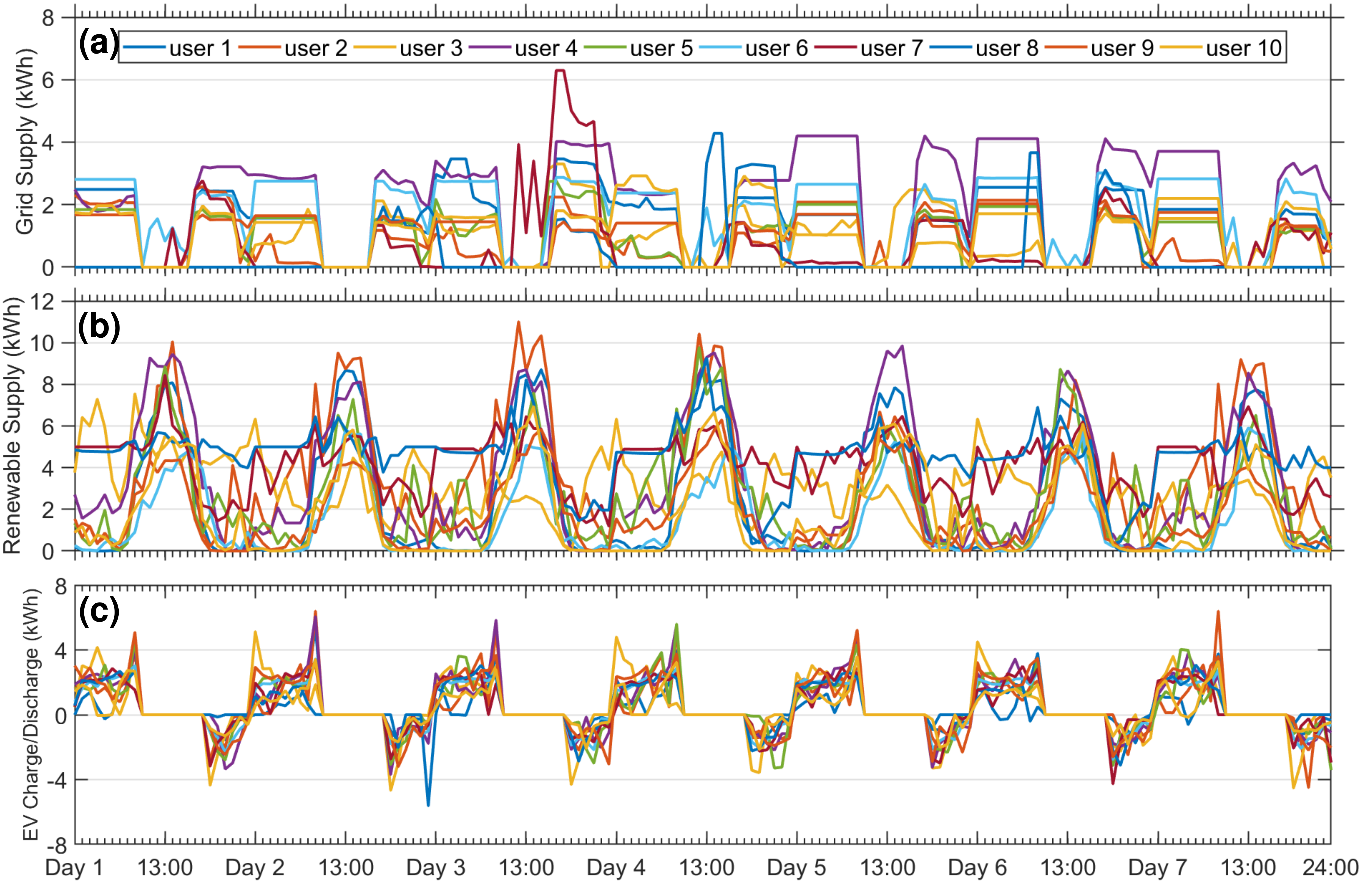}
    \caption{The optimal schedule of users for (a) grid energy purchase, (b) renewable energy supply, and (c) charge/discharge of the EV in the TEM scenario. In (c) positive value means charging and negative value means discharging.}
    \label{f:supply}
\end{figure}

\subsubsection{Power Scheduling in Smart Homes} Fig.~\ref{f:supply} shows the hourly time-series results of the optimized decisions of 10 users over one week (September 6-12, 2017) in the TEM scenario. We see the users enjoy local renewable generation, e.g., PV energy in the daytime, to serve their demand, and also purchase electricity from the grid mostly in the early morning and at night. Users' EVs perform V2H to discharge energy upon arrival home, as the evening is often the peak time for the residential load. Later at night, the EVs are charged to satisfy the charging demand before departure in the next morning.

\begin{figure}[!t]
    \centering
    \includegraphics[width=8.7cm]{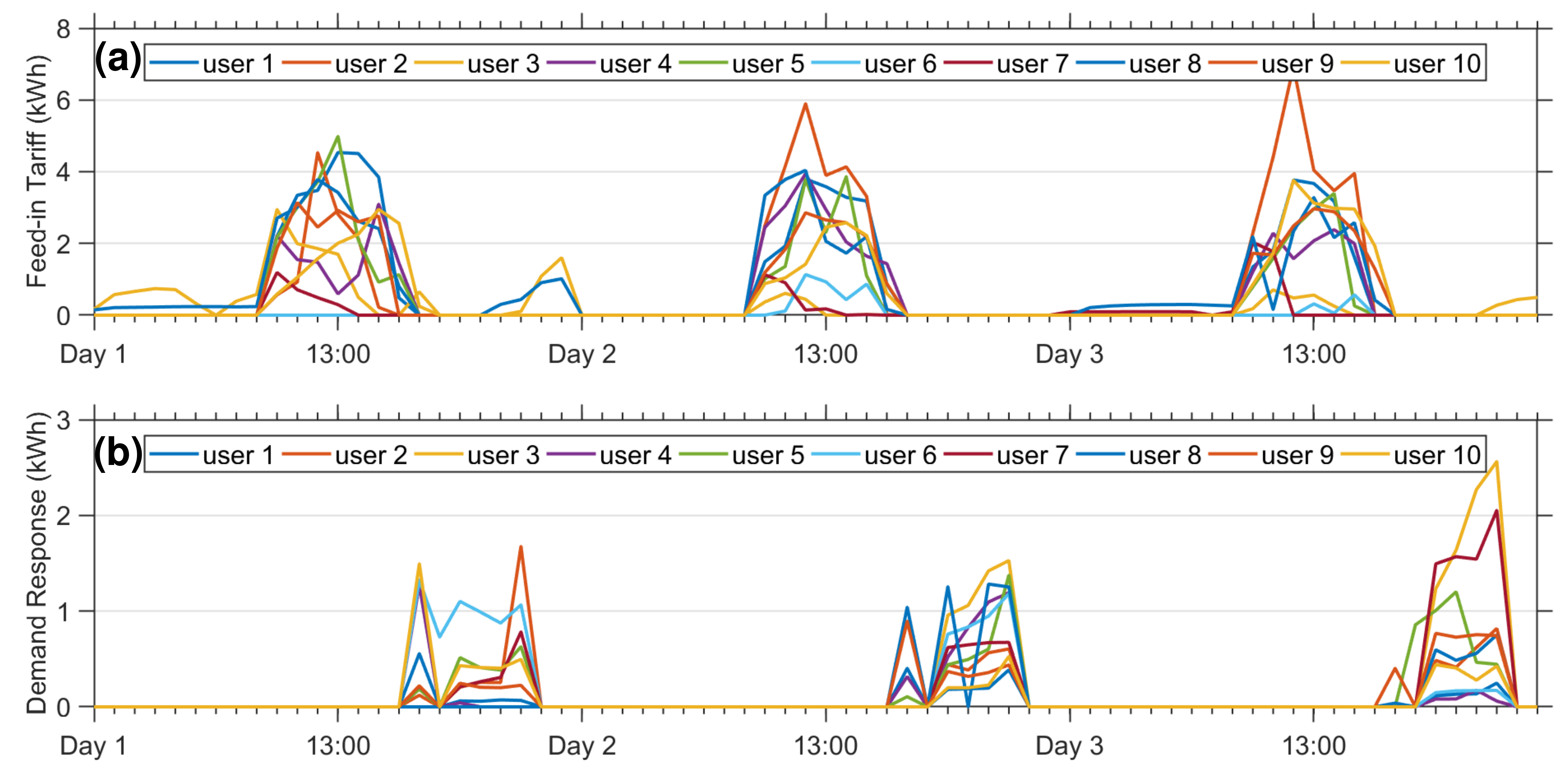}
    \caption{The optimal vertical transactive energy of users for (a) feed-in PV energy and (b) demand response in the TEM scenario.}
    \label{f:fitdr}
\end{figure}
\begin{figure}[!t]
    \centering
    \includegraphics[width=8.8cm]{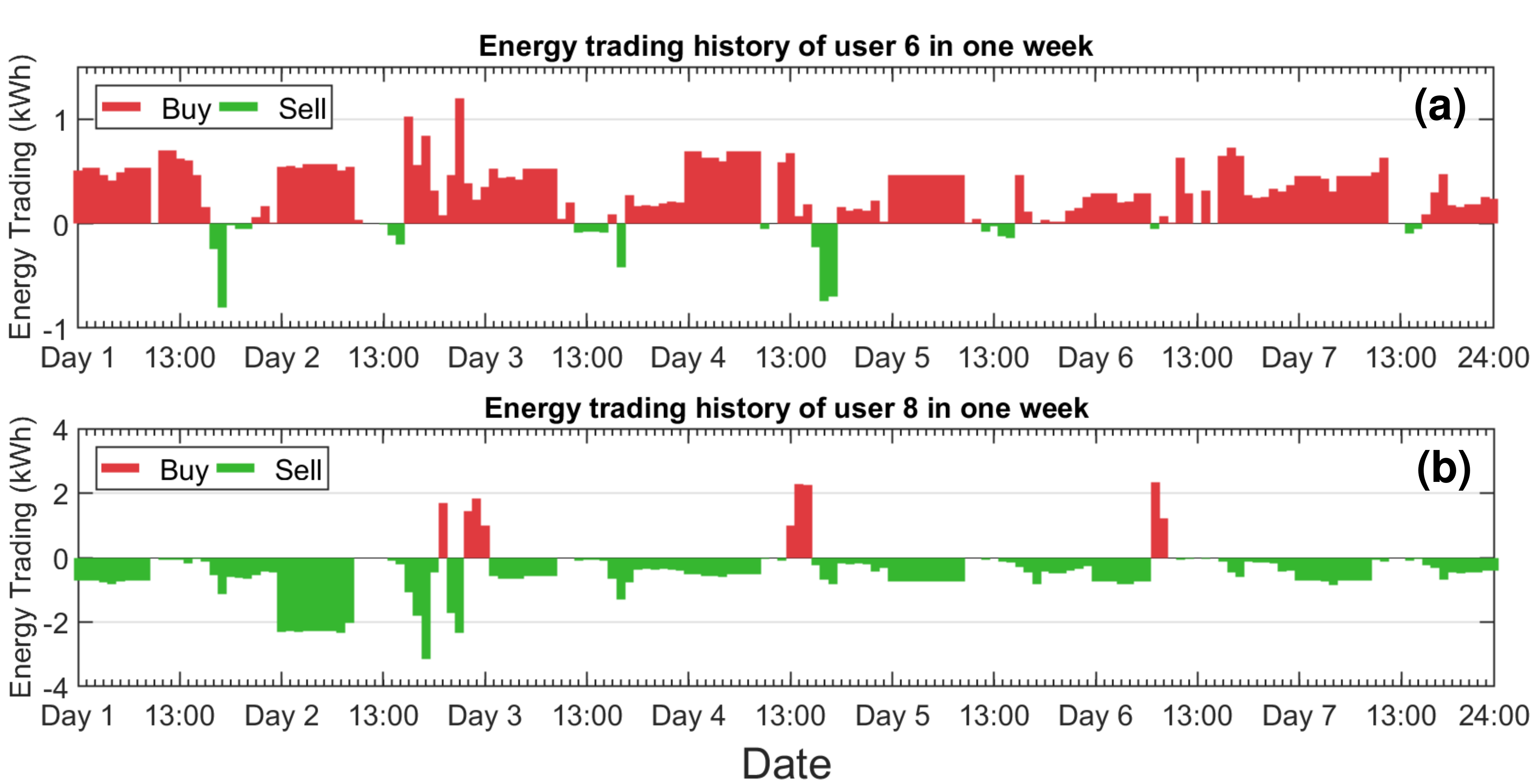}
    \caption{The optimal energy trading (i.e., horizontal transactive energy) of two typical users (a) user \#6 (b) user \#8.}
    \label{f:trading}
\end{figure}

\subsubsection{Transactive Energy within the Grid and Among Smart Homes} We plot the optimal decisions of users' vertical transactive energy in Fig.~\ref{f:fitdr} and horizontal transactive energy (i.e., energy trading) in Fig.~\ref{f:trading}. We can see from Fig.~\ref{f:fitdr} that users actively participated in the feed-in PV energy and demand response programs. They sell their extra PV energy back to the grid or trade with other smart homes. Users also provide demand response services to the grid during the peak hours, given that their internal power scheduling is jointly optimized. In Fig.~\ref{f:trading}, we can see two typical users: User \#6 is often in short of energy supply and thus trade to buy more energy from other users through the horizontal transactive energy system. User \# 8 is the opposite type with more local renewable energy generation. Thus, it sells more energy to help other users and gain some benefits through the horizontal transactive energy system. We can see that users actively participate in both vertical and horizontal transactive energy.

\begin{figure}[!t]
    \centering
    \includegraphics[width=8.7cm]{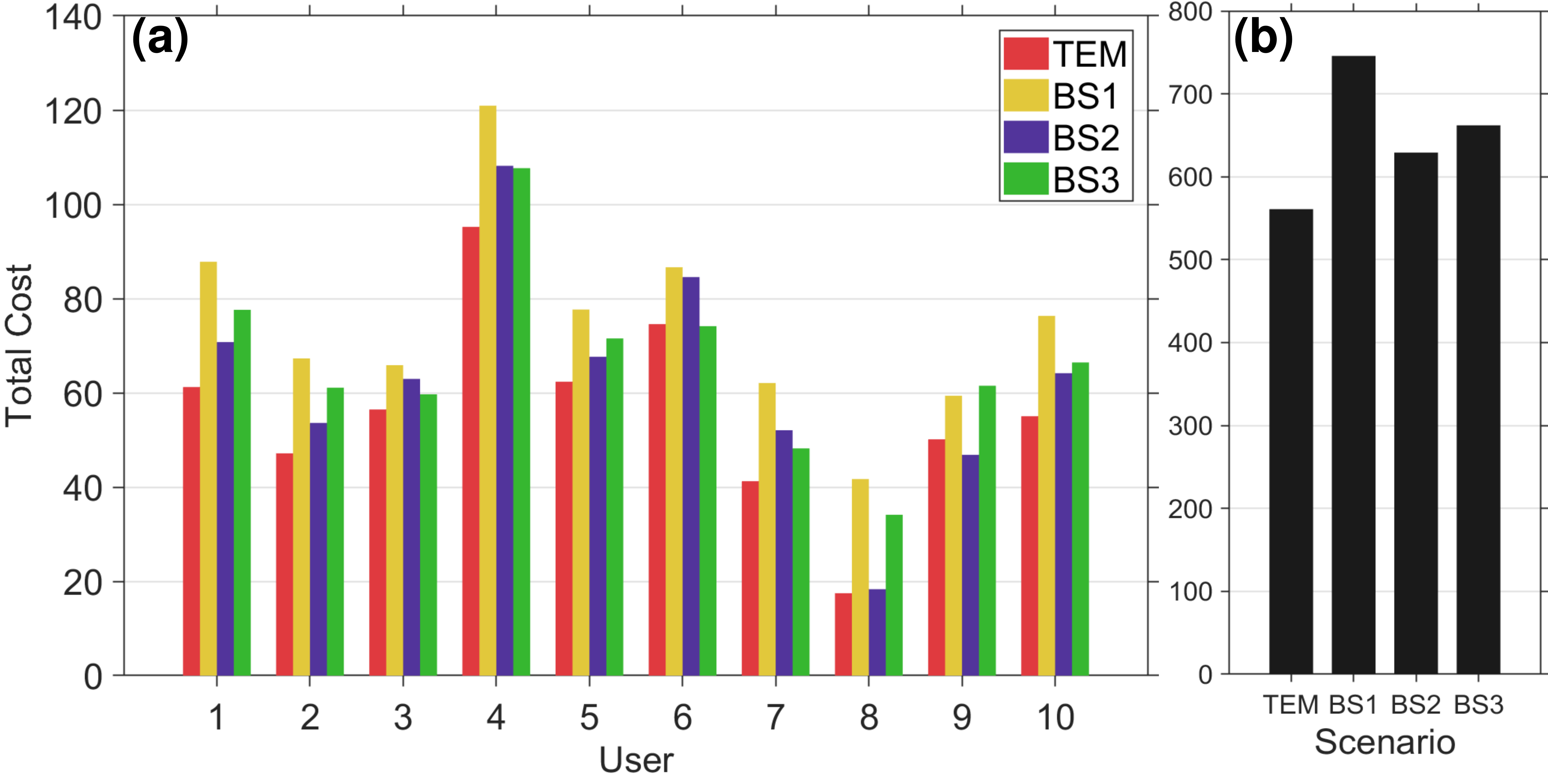}
    \caption{(a) The total costs of the ten users in one week. We compare the costs of the IoT blockchain-based transactive energy management (TEM) with three benchmark scenarios (BS1, BS2, and BS3). (b) Comparing the total costs of TEM with the benchmark scenarios.}
    \label{f:overall_cost}
\end{figure}

\subsubsection{Benefits of Transactive Energy}
We evaluate the benefits that transactive energy management brings to the users and compare the total costs of users in Fig.~\ref{f:overall_cost}. We compare our transactive energy management with three benchmarks introduced in Section~\ref{sec:benchmark}. Users in Benchmark Scenario 1 have the highest total costs as they do not participate in any transactive energy systems. Users can reduce their costs by performing either vertical transactive energy in Benchmark Scenario 2 or horizontal transactive energy in Benchmark Scenario 3. From Fig.~\ref{f:overall_cost}(a), we can see that the benefits of vertical transactive energy and horizontal transactive energy can be different from user to user, given the diverse behaviors in users' supply and demand. Nevertheless, the overall cost comparison of all users in Fig.~\ref{f:overall_cost}(b) shows that the users reduce their costs by about 16\% and 11\% from vertical and horizontal transactive energy, respectively. The holistic transactive energy achieves the highest cost reduction of 25\%. The above results show that our developed transactive energy management brings the most benefits to users, and our blockchain-based system enables such an efficient and trustworthy design.

%% file: main.bbl
\begin{thebibliography}{10}
\providecommand{\url}[1]{#1}
\csname url@samestyle\endcsname
\providecommand{\newblock}{\relax}
\providecommand{\bibinfo}[2]{#2}
\providecommand{\BIBentrySTDinterwordspacing}{\spaceskip=0pt\relax}
\providecommand{\BIBentryALTinterwordstretchfactor}{4}
\providecommand{\BIBentryALTinterwordspacing}{\spaceskip=\fontdimen2\font plus
\BIBentryALTinterwordstretchfactor\fontdimen3\font minus
  \fontdimen4\font\relax}
\providecommand{\BIBforeignlanguage}[2]{{%
\expandafter\ifx\csname l@#1\endcsname\relax
\typeout{** WARNING: IEEEtran.bst: No hyphenation pattern has been}%
\typeout{** loaded for the language `#1'. Using the pattern for}%
\typeout{** the default language instead.}%
\else
\language=\csname l@#1\endcsname
\fi
#2}}
\providecommand{\BIBdecl}{\relax}
\BIBdecl

\bibitem{li2017smart}
Y.~Li, X.~Cheng, Y.~Cao, D.~Wang, and L.~Yang, ``Smart choice for the smart
  grid: Narrowband internet of things (nb-iot),'' \emph{IEEE Internet of Things
  Journal}, vol.~5, no.~3, pp. 1505--1515, 2017.

\bibitem{sun2015comprehensive}
Q.~Sun, H.~Li, Z.~Ma, C.~Wang, J.~Campillo, Q.~Zhang, F.~Wallin, and J.~Guo,
  ``A comprehensive review of smart energy meters in intelligent energy
  networks,'' \emph{IEEE Internet of Things Journal}, vol.~3, no.~4, pp.
  464--479, 2015.

\bibitem{moghaddam2018fog}
M.~H.~Y. Moghaddam and A.~Leon-Garcia, ``A fog-based internet of energy
  architecture for transactive energy management systems,'' \emph{IEEE Internet
  of Things Journal}, vol.~5, no.~2, pp. 1055--1069, 2018.

\bibitem{mollah2020blockchain}
M.~B. Mollah, J.~Zhao, D.~Niyato, K.-Y. Lam, X.~Zhang, A.~M. Ghias, L.~H. Koh,
  and L.~Yang, ``Blockchain for future smart grid: A comprehensive survey,''
  \emph{IEEE Internet of Things Journal}, 2020.

\bibitem{nakamoto2008bitcoin}
\BIBentryALTinterwordspacing
S.~Nakamoto, ``Bitcoin: A peer-to-peer electronic cash system,'' White Paper,
  Bitcoin, 2008. [Online]. Available: \url{https://bitcoin.org/bitcoin.pdf}
\BIBentrySTDinterwordspacing

\bibitem{eth}
\BIBentryALTinterwordspacing
G.~Wood, ``Ethereum: A secure decentralised generalised transaction ledger,''
  Yellow Paper, Ethereum, 2019. [Online]. Available:
  \url{https://ethereum.github.io/yellowpaper/paper.pdf}
\BIBentrySTDinterwordspacing

\bibitem{dai2019blockchain}
H.-N. Dai, Z.~Zheng, and Y.~Zhang, ``Blockchain for internet of things: A
  survey,'' \emph{IEEE Internet of Things Journal}, vol.~6, no.~5, pp.
  8076--8094, 2019.

\bibitem{gai2019privacy}
K.~Gai, Y.~Wu, L.~Zhu, M.~Qiu, and M.~Shen, ``Privacy-preserving energy trading
  using consortium blockchain in smart grid,'' \emph{IEEE Trans. Ind.
  Informat.}, vol.~15, no.~6, pp. 3548--3558, 2019.

\bibitem{li2017consortium}
Z.~Li, J.~Kang, R.~Yu, D.~Ye, Q.~Deng, and Y.~Zhang, ``Consortium blockchain
  for secure energy trading in industrial internet of things,'' \emph{IEEE
  Trans. Ind. Informat.}, vol.~14, no.~8, pp. 3690--3700, 2017.

\bibitem{wan2019blockchain}
J.~Wan, J.~Li, M.~Imran, D.~Li \emph{et~al.}, ``A blockchain-based solution for
  enhancing security and privacy in smart factory,'' \emph{IEEE Trans. Ind.
  Informat.}, vol.~15, no.~6, pp. 3652--3660, 2019.

\bibitem{li2019blockchain}
Z.~Li, S.~Bahramirad \emph{et~al.}, ``Blockchain for decentralized transactive
  energy management system in networked microgrids,'' \emph{Elsevier Electr.
  J.}, vol.~32, no.~4, pp. 58--72, 2019.

\bibitem{gai2019permissioned}
K.~Gai, Y.~Wu, L.~Zhu, L.~Xu, and Y.~Zhang, ``Permissioned blockchain and edge
  computing empowered privacy-preserving smart grid networks,'' \emph{IEEE
  Internet of Things Journal}, vol.~6, no.~5, pp. 7992--8004, 2019.

\bibitem{yang2020blockchain}
Q.~Yang and H.~Wang, ``Blockchain-empowered socially optimal transactive energy
  system: Framework and implementation,'' \emph{IEEE Transactions on Industrial
  Informatics}, pp. 1--11, 2020.

\bibitem{aitzhan2016security}
N.~Z. Aitzhan and D.~Svetinovic, ``Security and privacy in decentralized energy
  trading through multi-signatures, blockchain and anonymous messaging
  streams,'' \emph{IEEE Trans. Dependable Secure Comput.}, vol.~15, no.~5, pp.
  840--852, 2016.

\bibitem{mihaylov2014nrgcoin}
M.~Mihaylov, S.~Jurado, N.~Avellana, K.~Van~Moffaert, I.~M. de~Abril, and
  A.~Now{\'e}, ``{NRG}coin: Virtual currency for trading of renewable energy in
  smart grids,'' in \emph{11th International conference on the European energy
  market (EEM14)}.\hskip 1em plus 0.5em minus 0.4em\relax IEEE, 2014, pp. 1--6.

\bibitem{exergy}
\BIBentryALTinterwordspacing
``Building a robust value mechanism to facilitate transactive energy,''
  Technical Whitepaper, {LO3 Energy}, 2017. [Online]. Available:
  \url{https://exergy.energy/wp-content/uploads/2017/12/Exergy-Whitepaper-v8.pdf}
\BIBentrySTDinterwordspacing

\bibitem{wang2019energy}
S.~Wang, A.~F. Taha, J.~Wang, K.~Kvaternik, and A.~Hahn, ``Energy crowdsourcing
  and peer-to-peer energy trading in blockchain-enabled smart grids,''
  \emph{IEEE Transactions on Systems, Man, and Cybernetics Systems}, vol.~49,
  no.~8, pp. 1612--1623, 2019.

\bibitem{sabounchi2017towards}
M.~Sabounchi and J.~Wei, ``Towards resilient networked microgrids:
  Blockchain-enabled peer-to-peer electricity trading mechanism,'' in
  \emph{Proc. IEEE Conference on Energy Internet and Energy System
  Integration}.\hskip 1em plus 0.5em minus 0.4em\relax IEEE, 2017, pp. 1--5.

\bibitem{wang2019bbars}
H.~Wang, Q.~Wang, D.~He, Q.~Li, and Z.~Liu, ``Bbars: Blockchain-based anonymous
  rewarding scheme for v2g networks,'' \emph{IEEE Internet of Things Journal},
  vol.~6, no.~2, pp. 3676--3687, 2019.

\bibitem{thomas2019general}
L.~Thomas, Y.~Zhou, C.~Long, J.~Wu, and N.~Jenkins, ``A general form of smart
  contract for decentralized energy systems management,'' \emph{Nature Energy},
  vol.~4, no.~2, pp. 140--149, 2019.

\bibitem{iota}
\BIBentryALTinterwordspacing
S.~Popov, ``The tangle,'' Internet Draft, IOTA, 2018. [Online]. Available:
  \url{https://assets.ctfassets.net/r1dr6vzfxhev/2t4uxvsIqk0EUau6g2sw0g/45eae33637ca92f85dd9f4a3a218e1ec/iota1_4_3.pdf}
\BIBentrySTDinterwordspacing

\bibitem{danzi2019delay}
P.~Danzi, A.~E. Kal{\o}r, {\v{C}}.~Stefanovi{\'c}, and P.~Popovski, ``Delay and
  communication tradeoffs for blockchain systems with lightweight iot
  clients,'' \emph{IEEE Internet of Things Journal}, vol.~6, no.~2, pp.
  2354--2365, 2019.

\bibitem{zheng2018user}
S.~Zheng, N.~Apthorpe, M.~Chetty, and N.~Feamster, ``User perceptions of smart
  home iot privacy,'' \emph{Proceedings of the ACM on Human-Computer
  Interaction}, vol.~2, no. CSCW, pp. 1--20, 2018.

\bibitem{cha2018privacy}
S.-C. Cha, T.-Y. Hsu, Y.~Xiang, and K.-H. Yeh, ``Privacy enhancing technologies
  in the internet of things: Perspectives and challenges,'' \emph{IEEE Internet
  of Things Journal}, vol.~6, no.~2, pp. 2159--2187, 2018.

\bibitem{pnnl}
J.~Lian \emph{et~al.}, ``Transactive system, part {I}: Theoretical
  underpinnings of payoff functions, control decisions, information privacy,
  and solution concepts,'' Pacific Northwest National Laboratory, Tech. Rep.
  PNNL-27235, December 2017.

\bibitem{zhang2019security}
R.~Zhang, R.~Xue, and L.~Liu, ``Security and privacy on blockchain,'' \emph{ACM
  Computing Surveys (CSUR)}, vol.~52, no.~3, pp. 1--34, 2019.

\bibitem{dorri2017blockchain}
A.~Dorri, S.~S. Kanhere, R.~Jurdak, and P.~Gauravaram, ``Blockchain for iot
  security and privacy: The case study of a smart home,'' in \emph{2017 IEEE
  international conference on pervasive computing and communications workshops
  (PerCom workshops)}.\hskip 1em plus 0.5em minus 0.4em\relax IEEE, 2017, pp.
  618--623.

\bibitem{ferrag2018blockchain}
M.~A. Ferrag, M.~Derdour, M.~Mukherjee, A.~Derhab, L.~Maglaras, and H.~Janicke,
  ``Blockchain technologies for the internet of things: Research issues and
  challenges,'' \emph{IEEE Internet of Things Journal}, vol.~6, no.~2, pp.
  2188--2204, 2018.

\bibitem{cui2019}
S.~Cui, Y.-W. Wang, and J.-W. Xiao, ``Peer-to-peer energy sharing among smart
  energy buildings by distributed transaction,'' \emph{IEEE Trans. Smart Grid},
  vol.~10, no.~6, pp. 6491--6501, 2019.

\bibitem{bitcoinfull}
\BIBentryALTinterwordspacing
\emph{Running A Full Node: Support the Bitcoin network by running your own full
  node}, Bitcoin core, accessed Aug. 1, 2020. [Online]. Available:
  \url{https://bitcoin.org/en/full-node#what-is-a-full-node}
\BIBentrySTDinterwordspacing

\bibitem{wenbo2018survey}
\BIBentryALTinterwordspacing
W.~Wang, D.~T. Hoang, Z.~Xiong, D.~Niyato, P.~Wang, P.~Hu, and Y.~Wen, ``A
  survey on consensus mechanisms and mining management in blockchain
  networks,'' \emph{CoRR}, 2018. [Online]. Available:
  \url{http://arxiv.org/abs/1805.02707}
\BIBentrySTDinterwordspacing

\bibitem{natoli2019deconstructing}
C.~Natoli, J.~Yu, V.~Gramoli, and P.~Esteves-Verissimo, ``Deconstructing
  blockchains: A comprehensive survey on consensus, membership and structure,''
  \emph{arXiv preprint, arXiv:1908.08316}, 2019.

\bibitem{castro1999practical}
M.~Castro and B.~Liskov, ``Practical {B}yzantine fault tolerance,'' in
  \emph{Proc. OSDI '99}, New Orleans, USA, 1999, pp. 173--186.

\bibitem{boyd2011distributed}
S.~Boyd, N.~Parikh \emph{et~al.}, ``Distributed optimization and statistical
  learning via the alternating direction method of multipliers,'' \emph{Found.
  Trends Mach. Learn.}, vol.~3, no.~1, pp. 1--122, 2011.

\bibitem{wang2016incentivizing}
H.~Wang and J.~Huang, ``Incentivizing energy trading for interconnected
  microgrids,'' \emph{IEEE Transactions on Smart Grid}, vol.~9, no.~4, pp.
  2647--2657, 2018.

\bibitem{rpi}
\BIBentryALTinterwordspacing
\emph{Raspberry {Pi} 3 Model {B}+}, Website Documentation, Raspberry Pi
  Foundation, accessed Dec. 28, 2019. [Online]. Available:
  \url{https://www.raspberrypi.org/products/raspberry-pi-3-model-b-plus/}
\BIBentrySTDinterwordspacing

\bibitem{quorum}
\BIBentryALTinterwordspacing
\emph{Quorum}, J.P. Morgan Chase, accessed Oct. 1, 2019, release v2.3.0.
  [Online]. Available: \url{https://github.com/jpmorganchase/quorum}
\BIBentrySTDinterwordspacing

\bibitem{wang2015joint}
H.~Wang and J.~Huang, ``Joint investment and operation of microgrid,''
  \emph{IEEE Transactions on Smart Grid}, vol.~8, no.~2, pp. 833--845, 2017.

\bibitem{pecan}
\BIBentryALTinterwordspacing
\emph{Energy Research}, Website, Pecan Street Inc., accessed Oct. 1, 2019.
  [Online]. Available: \url{https://www.pecanstreet.org/dataport/}
\BIBentrySTDinterwordspacing

\bibitem{WebAssembly}
\BIBentryALTinterwordspacing
\emph{WebAssembly}, Website. [Online]. Available: \url{http://webassembly.org}
\BIBentrySTDinterwordspacing

\end{thebibliography}
